\documentclass[aps,prl,superscriptaddress,showpacs,byrevtex]{revtex4}

\usepackage{graphicx} 
\usepackage{dcolumn}  

\graphicspath{{ps}}

\newcommand{\BF}{\ensuremath{{\cal{B}}}}

\begin{document}

\vbox{ \hbox{   }
                 \hbox{\hspace*{140mm}BELLE-CONF-0225}  
                 \hbox{\hspace*{140mm}Parallel Session 8}
                 \hbox{\hspace*{140mm}ABS713}
}

\title{\quad\\[0.5cm] Study of charmless $B$ decays to three-kaon final states}

\affiliation{Aomori University, Aomori}
\affiliation{Budker Institute of Nuclear Physics, Novosibirsk}
\affiliation{Chiba University, Chiba}
\affiliation{Chuo University, Tokyo}
\affiliation{University of Cincinnati, Cincinnati OH}
\affiliation{University of Frankfurt, Frankfurt}
\affiliation{Gyeongsang National University, Chinju}
\affiliation{University of Hawaii, Honolulu HI}
\affiliation{High Energy Accelerator Research Organization (KEK), Tsukuba}
\affiliation{Hiroshima Institute of Technology, Hiroshima}
\affiliation{Institute of High Energy Physics, Chinese Academy of Sciences, Beijing}
\affiliation{Institute of High Energy Physics, Vienna}
\affiliation{Institute for Theoretical and Experimental Physics, Moscow}
\affiliation{J. Stefan Institute, Ljubljana}
\affiliation{Kanagawa University, Yokohama}
\affiliation{Korea University, Seoul}
\affiliation{Kyoto University, Kyoto}
\affiliation{Kyungpook National University, Taegu}
\affiliation{Institut de Physique des Hautes \'Energies, Universit\'e de Lausanne, Lausanne}
\affiliation{University of Ljubljana, Ljubljana}
\affiliation{University of Maribor, Maribor}
\affiliation{University of Melbourne, Victoria}
\affiliation{Nagoya University, Nagoya}
\affiliation{Nara Women's University, Nara}
\affiliation{National Kaohsiung Normal University, Kaohsiung}
\affiliation{National Lien-Ho Institute of Technology, Miao Li}
\affiliation{National Taiwan University, Taipei}
\affiliation{H. Niewodniczanski Institute of Nuclear Physics, Krakow}
\affiliation{Nihon Dental College, Niigata}
\affiliation{Niigata University, Niigata}
\affiliation{Osaka City University, Osaka}
\affiliation{Osaka University, Osaka}
\affiliation{Panjab University, Chandigarh}
\affiliation{Peking University, Beijing}
\affiliation{Princeton University, Princeton NJ}
\affiliation{RIKEN BNL Research Center, Brookhaven NY}
\affiliation{Saga University, Saga}
\affiliation{University of Science and Technology of China, Hefei}
\affiliation{Seoul National University, Seoul}
\affiliation{Sungkyunkwan University, Suwon}
\affiliation{University of Sydney, Sydney NSW}
\affiliation{Tata Institute of Fundamental Research, Bombay}
\affiliation{Toho University, Funabashi}
\affiliation{Tohoku Gakuin University, Tagajo}
\affiliation{Tohoku University, Sendai}
\affiliation{University of Tokyo, Tokyo}
\affiliation{Tokyo Institute of Technology, Tokyo}
\affiliation{Tokyo Metropolitan University, Tokyo}
\affiliation{Tokyo University of Agriculture and Technology, Tokyo}
\affiliation{Toyama National College of Maritime Technology, Toyama}
\affiliation{University of Tsukuba, Tsukuba}
\affiliation{Utkal University, Bhubaneswer}
\affiliation{Virginia Polytechnic Institute and State University, Blacksburg VA}
\affiliation{Yokkaichi University, Yokkaichi}
\affiliation{Yonsei University, Seoul}
  \author{K.~Abe}\affiliation{High Energy Accelerator Research Organization (KEK), Tsukuba} 
  \author{K.~Abe}\affiliation{Tohoku Gakuin University, Tagajo} 
  \author{N.~Abe}\affiliation{Tokyo Institute of Technology, Tokyo} 
  \author{R.~Abe}\affiliation{Niigata University, Niigata} 
  \author{T.~Abe}\affiliation{Tohoku University, Sendai} 
  \author{I.~Adachi}\affiliation{High Energy Accelerator Research Organization (KEK), Tsukuba} 
  \author{Byoung~Sup~Ahn}\affiliation{Korea University, Seoul} 
  \author{H.~Aihara}\affiliation{University of Tokyo, Tokyo} 
  \author{M.~Akatsu}\affiliation{Nagoya University, Nagoya} 
  \author{M.~Asai}\affiliation{Hiroshima Institute of Technology, Hiroshima} 
% ...
\collaboration{The Belle Collaboration}

\noaffiliation

\begin{abstract}
  We report on a study of charmless $B$ meson decays to three-kaon final 
states. The results are obtained with a 78.7\,fb$^{-1}$ data sample collected
on the $\Upsilon(4S)$ resonance by the Belle detector operating at the KEKB 
asymmetric energy $e^+e^-$ collider. The branching fractions for $B$ decays to 
three-body $K^+K^+K^-$, $K^0K^+K^-$, $K_SK_SK^+$, and $K_SK_SK_S$ final states
are presented. We also make a first attempt to perform an isospin analysis of
the three-kaon final states.
\end{abstract}

\maketitle

\tighten

{\renewcommand{\thefootnote}{\fnsymbol{footnote}}}
\setcounter{footnote}{0}


\section{Introduction}

  Studies of three-body $B$ decays can significantly broaden the understanding of
$B$ meson decay mechanisms and provide additional possibilities for CP violation
searches. Analysis of the $K\pi\pi$ and $KK\pi$ final states is presented 
in Refs.~\cite{bc226,b2khh}. In this paper we report results on the study of
charmless $B$ meson decays to three-kaon final states. The analysis is based
on a 78.7\,fb$^{-1}$ data sample, which contains 85.0 million $B\bar{B}$ pairs,
collected  with the Belle detector  operating at the KEKB 
asymmetric-energy $e^+e^-$ (3.5 on 8~GeV) collider~\cite{KEKB} with a 
center-of-mass energy at the $\Upsilon(4S)$ resonance. All results reported here
are preliminary.

\section{The Belle detector}

  The Belle detector~\cite{Belle} is a large-solid-angle magnetic spectrometer
that consists
of a three-layer silicon vertex detector (SVD), a 50-layer central drift chamber
(CDC) for charged particle tracking and specific ionization measurement 
($dE/dx$), an array of aerogel threshold \v{C}erenkov counters (ACC), 
time-of-flight scintillation counters (TOF), and an array of 8736 CsI(Tl) 
crystals for electromagnetic calorimetry (ECL) located inside a superconducting
solenoid coil that provides a 1.5~T magnetic field. An iron flux return located
outside the coil is instrumented to detect $K_L$ mesons and to identify muons
(KLM). Electron identification is based on a combination of CDC $dE/dx$ 
measurements, the response of the ACC, and the position, shape and energy 
deposition of the associated ECL shower.

  Charged hadron identification is accomplished by combining the 
responses of the ACC and the TOF with $dE/dx$ measurements in the CDC into a 
single value using the likelihood method:
\[ {\cal{L}}(h) = 
{\cal{L}}^{ACC}(h)\times {\cal{L}}^{TOF}(h)\times{\cal{L}}^{CDC}(h),\] where $h$
stands for the hadron type ($p$, $K$, $\pi$). Charged tracks are identified
as protons, pions or kaons by imposing requirements on the likelihood ratios 
(PID):
\[{\rm PID}(p)=\frac{{\cal{L}}(p)}{{\cal{L}}(p)+{\cal{L}}(K)};~~
{\rm PID}(K)=\frac{{\cal{L}}(K)}{{\cal{L}}(K)+{\cal{L}}(\pi)};~~
{\rm PID}(\pi)=
\frac{{\cal{L}}(\pi)}{{\cal{L}}(K)+{\cal{L}}(\pi)}=1-{\rm PID}(K)\]
  At large momenta ($>$2.5 GeV/$c$) only the ACC and $dE/dx$ are used since the
TOF provides no significant separation of kaons and pions.
 We use a GEANT based Monte Carlo (MC) simulation to model the response of the
detector and determine acceptance~\cite{sim}.

\section{Event selection}

  The selection criteria are similar to those used in the analysis of $B$
decays to the $K\pi\pi$ and $KK\pi$ final states~\cite{bc226}. Charged tracks
are selected with a set of track quality requirements based on the average hit
residual and on the distances of closest approach to the interaction point in
the plane perpendicular to the beam and the plane containing the beam and the
track. We also require that the transverse track momenta be greater than 
0.1~GeV/$c$ to reduce the low momentum combinatorial background. For charged 
kaon identification we impose a requirement on PID($K$), which has 86\% 
efficiency and a 7\% fake rate from misidentified pions. Charged tracks that 
are positively identified as electrons or protons are excluded. Since the muon
identification efficiency and fake rate vary significantly with the track 
momentum, we do not veto muons to avoid additional systematic errors.

  Neutral kaons are reconstructed via the decay chain
$K^0(\bar{K}^0)\to K_S\to\pi^+\pi^-$. The invariant mass of the two pions is 
required to be in the range $|M(\pi^+\pi^-)-M_{K^0}|<12$~MeV/$c^2$. 
The displacement of the $\pi^+\pi^-$ vertex from the interaction point (IP) in
the transverse ($r$-$\phi$) plane is required to be greater than 0.1~cm and 
less than 20~cm.
The direction of the combined pion pair momentum in the $r$-$\phi$ plane is 
required to be within 0.2 rad of the direction from the IP to the displaced 
vertex.

  We reconstruct $B$ mesons in the $K^+K^+K^-$, $K_SK^+K^-$, $K_SK_SK^+$ and 
$K_SK_SK_S$ three-body final states. The inclusion of the charge conjugate mode
is implied throughout this report. The candidate events are identified by their
center-of-mass (c.m.) energy difference, $\Delta E=(\sum_iE_i)-E_{\rm b}$, and 
the beam constrained mass, $M_{\rm bc}=\sqrt{E^2_{\rm b}-(\sum_i\vec{p}_i)^2}$,
where $E_{\rm b}=\sqrt{s}/2$ is the beam energy in the c.m.\ frame, and 
$\vec{p}_i$ and $E_i$ are the c.m.\ three-momenta and energies of the candidate
$B$ meson decay
products. In the first stage, we select events with $M_{\rm bc}>5.20$~GeV/$c^2$
and $-0.30<\Delta E<0.50$~GeV, which is a larger range than that used in our 
previous report~\cite{b2khh}. This allows for more detailed studies of the 
background. For subsequent analysis, we also define a {\it signal} region of 
$|M_{\rm bc}-M_B|<9$~MeV/$c^2$ and $|\Delta E|<0.04$~GeV and a $\Delta E$ 
{\it sideband} region defined as $0.05$~GeV $<|\Delta E|<0.15$~GeV.

  To determine the signal yield, we use events with $M_{\rm bc}$ in the signal
region and fit the $\Delta E$ distribution to the sum of a signal distribution
and an empirical background. The $\Delta E$ signal shape is parameterized by
the sum of two Gaussian functions with the same mean. The widths and the
relative fractions of the two Gaussians are determined from a MC simulation.
We find that the signal MC simulation gives, in general, a narrower $\Delta E$ 
width than data. To correct for this, we introduce a scale factor that is 
determined from the comparison of the $\Delta E$ widths for 
$B^+\to\bar{D}^0\pi^+$ events in MC and experimental data; this correction is
about 10\%. The background from $q\bar{q}$ continuum events is 
represented by a linear function. The $\Delta E$ shape of the $B\bar{B}$ 
background is determined from MC simulation, as described below.

\section{Background suppression}

  To suppress the combinatorial background from $e^+e^-\to q\bar{q}$ continuum
events, we use a set of variables that characterize the event topology. We 
require $|\cos\theta_{\rm thr}|<0.80$, where $\theta_{\rm thr}$ is the angle 
between the thrust axis of the $B$ candidate and that of the rest of the event;
the distribution of $|\cos\theta_{\rm thr}|$ is peaked near 1.0 for $q\bar{q}$
and is nearly flat for $B\bar{B}$ events. We also use a Fisher 
discriminant~\cite{fisher}, ${\cal{F}}$, formed from nine variables of a
``Virtual Calorimeter''~\cite{vcal}, the angle between the candidate thrust
axis and beam axis, and the angle between the $B$ candidate direction and beam
axis. We make a requirement on ${\cal{F}}$ that rejects 53\% of the remaining
$q\bar{q}$ background while retaining 89\% of the signal.

  We also consider backgrounds that come from
other $B$ decays. We conventionally subdivide this background into two types.
One type is the so called ``generic'' $B\bar{B}$ background that originates
from the dominant $b\to c$ tree transition. The description of these decays is
taken from an updated version of the CLEO group event generator~\cite{sim}.
We find that the dominant background from this source is due to $B\to Dh$ 
decays, where $h$ stands for a charged pion or kaon. To suppress this 
background, we reject events where any two-particle invariant mass is consistent
within 15~MeV (2.5$\sigma$) with $D^0\to K^+K^-$, $D^0\to K^-\pi^+$ or 
$D^+\to\bar{K}^0\pi^+$. We also reject events with a $K^+K^-$ invariant mass 
that is consistent with $\chi_{c0}\to K^+K^-$ 
($|M(K^+K^-)-M_{\chi_{c0}}|<50$~MeV). The other potential source of background
is rare charmless $B$ decays that proceed via $b\to s(d)$ penguins or $b\to u$ 
tree transitions. Since these final states are not included in the main 
generator decay table, they are generated separately. We studied a large set 
of potentially dangerous two-, three-, and four-body final states.
We do not find any rare charmless $B$ decay mode that produces a significant 
background to the three-kaon final states.


\begin{table}[t]
  \caption{Summary of results for $B$ meson decays to three-body 
           charmless hadronic final states. The branching fractions and
           90\% confedence level (CL) upper limits (UL) are quoted in units
           $10^{-6}$. For the modes with one neutral kaon the quoted 
           reconstruction efficiency includes the $K^0\to K_S\to \pi^+\pi^-$
           branching fraction. For modes with more than one neutral kaon only
           the $K_S\to \pi^+\pi^-$ branching fraction is taken into account.}
  \medskip
  \label{tab:defitall}
  \begin{tabular}{lccccc} \hline \hline
  Three-body~   &~~~Efficiency~~~&~~~Signal Yield~~~
                &~~~${\cal{B}}$ (90\% CL UL)~~~&~~Reference~~&~~Results from \\
  ~~~~mode      &      (\%)      &   (events)
                &                &    & Ref.~\cite{b2khh}     \\ \hline
$K^+K^+K^-$ & $23.5\pm0.50$ & $565\pm30$ & $33.0\pm1.8\pm3.2$  & This Work
                                         & $35.3\pm3.7\pm4.5$ \\
$K^0K^+K^-$ & $7.20\pm0.17$ & $149\pm15$ & $29.3\pm3.4\pm4.1$  &$-_{''}-$ & -- \\
$K_SK_SK^+$ & $6.78\pm0.19$ & $66.5\pm9.3$ & $13.4\pm1.9\pm1.5$&$-_{''}-$ & -- \\
$K_SK_SK_S$ & $3.98\pm0.17$ & $12.2^{+4.5}_{-3.8}$ 
                                  & $4.3^{+1.6}_{-1.4}\pm0.75$ &$-_{''}-$ & -- \\
\hline
$K^+K^-\pi^+$ & $13.8\pm0.31$ & $93.7\pm23.2$  & $9.3\pm2.3~(<13)$ & \cite{bc226}
                                               &     $<12$    \\
$K^0K^+\pi^-$ & $4.53\pm0.16$ & $26.8\pm16.6$  & $8.4\pm5.2~(<15)$ & \cite{bc226}
                                               &      --      \\
\hline
$\bar{D}^0\pi^+$, $\bar{D}^0\to K^+\pi^-$ & $28.8\pm0.57$ & $4000\pm66$ &        --         &  --  & -- \\
$D^-\pi^+$, $D^-\to K^0\pi^-$     & $8.24\pm0.14$ & $ 451\pm25$ &        --         &  --  & -- \\
\hline \hline
  \end{tabular}
\end{table}

\section{Results of the analysis}

\begin{figure}[t]
  \includegraphics[width=0.49\textwidth]{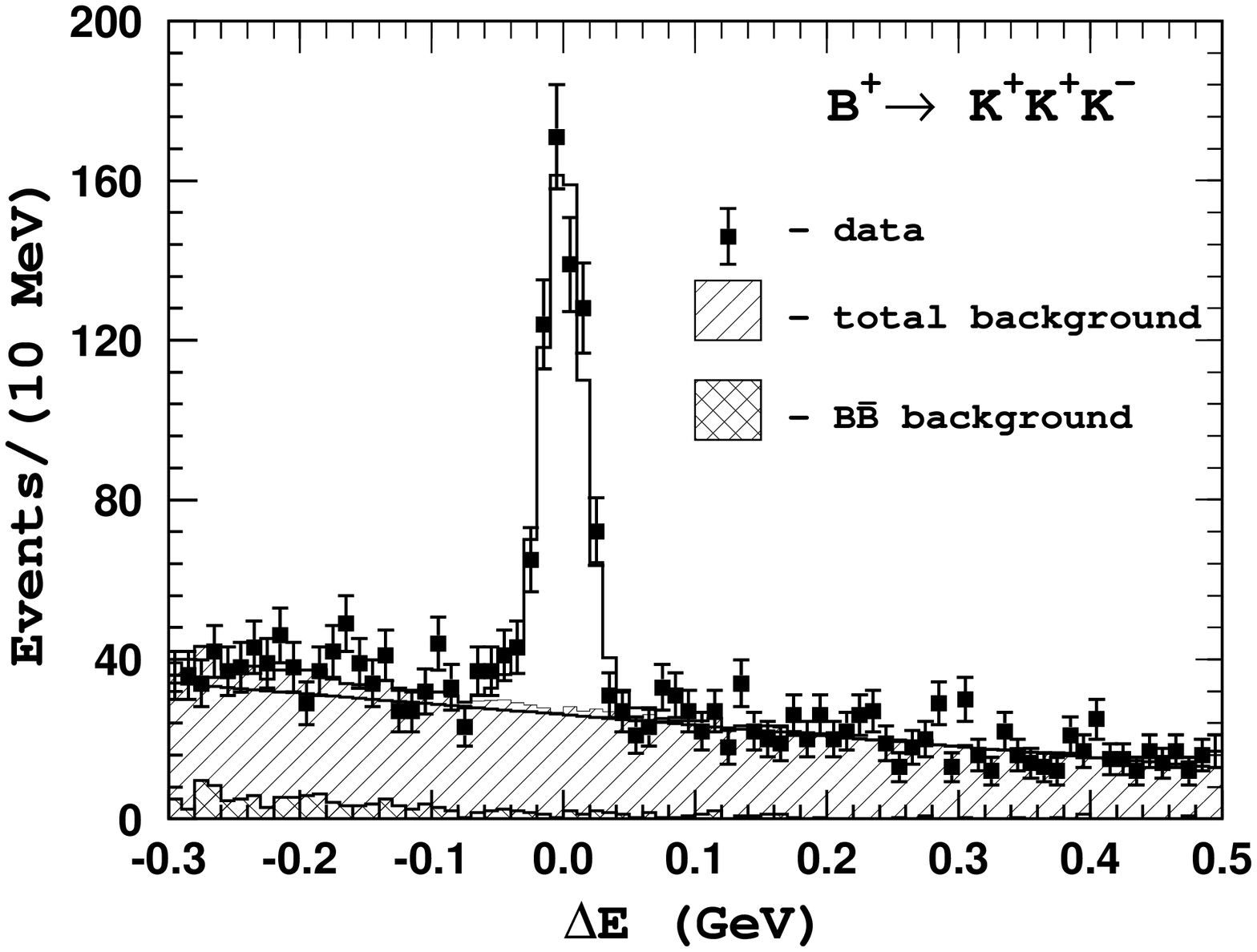} \hfill
  \includegraphics[width=0.49\textwidth]{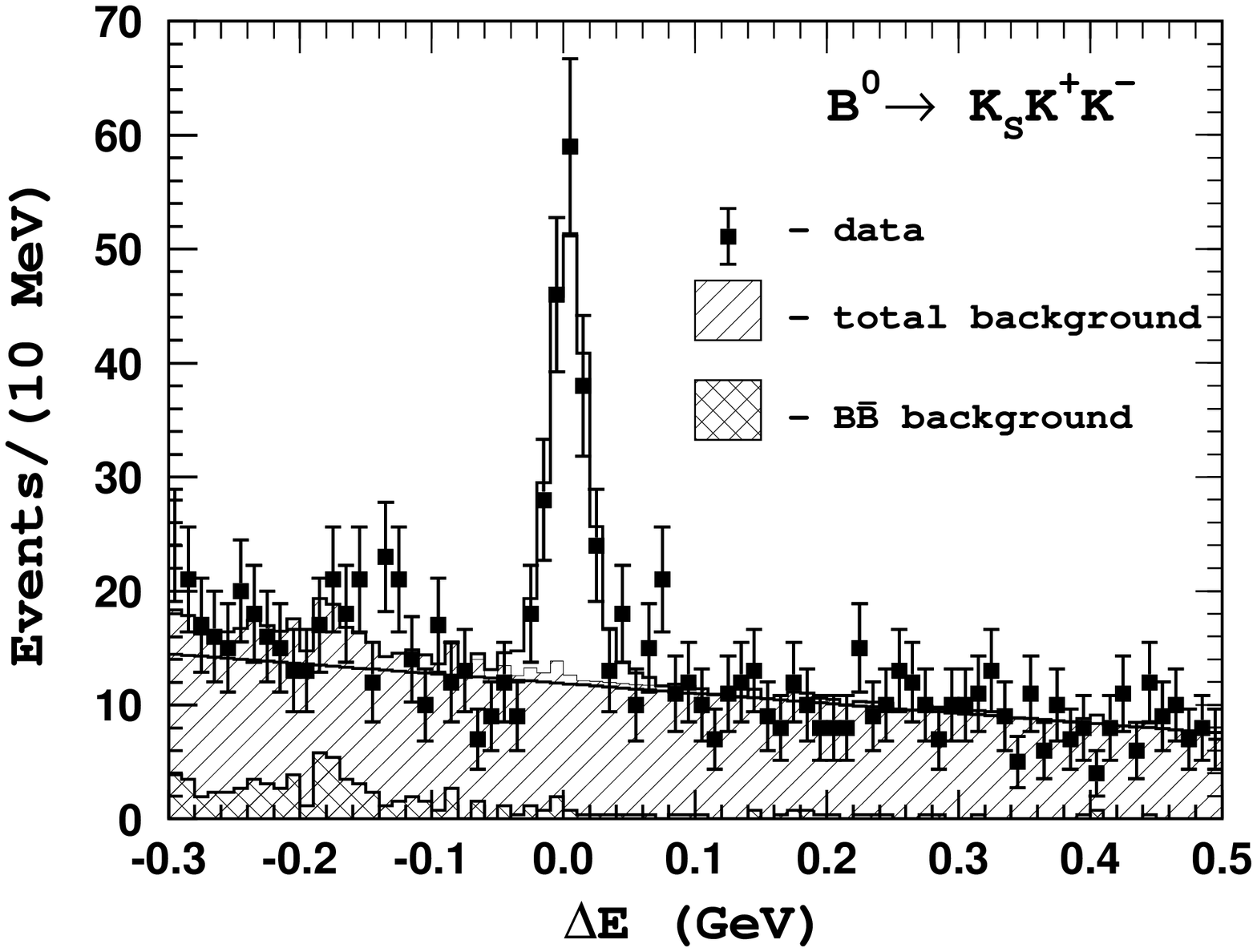}\\
  \includegraphics[width=0.49\textwidth]{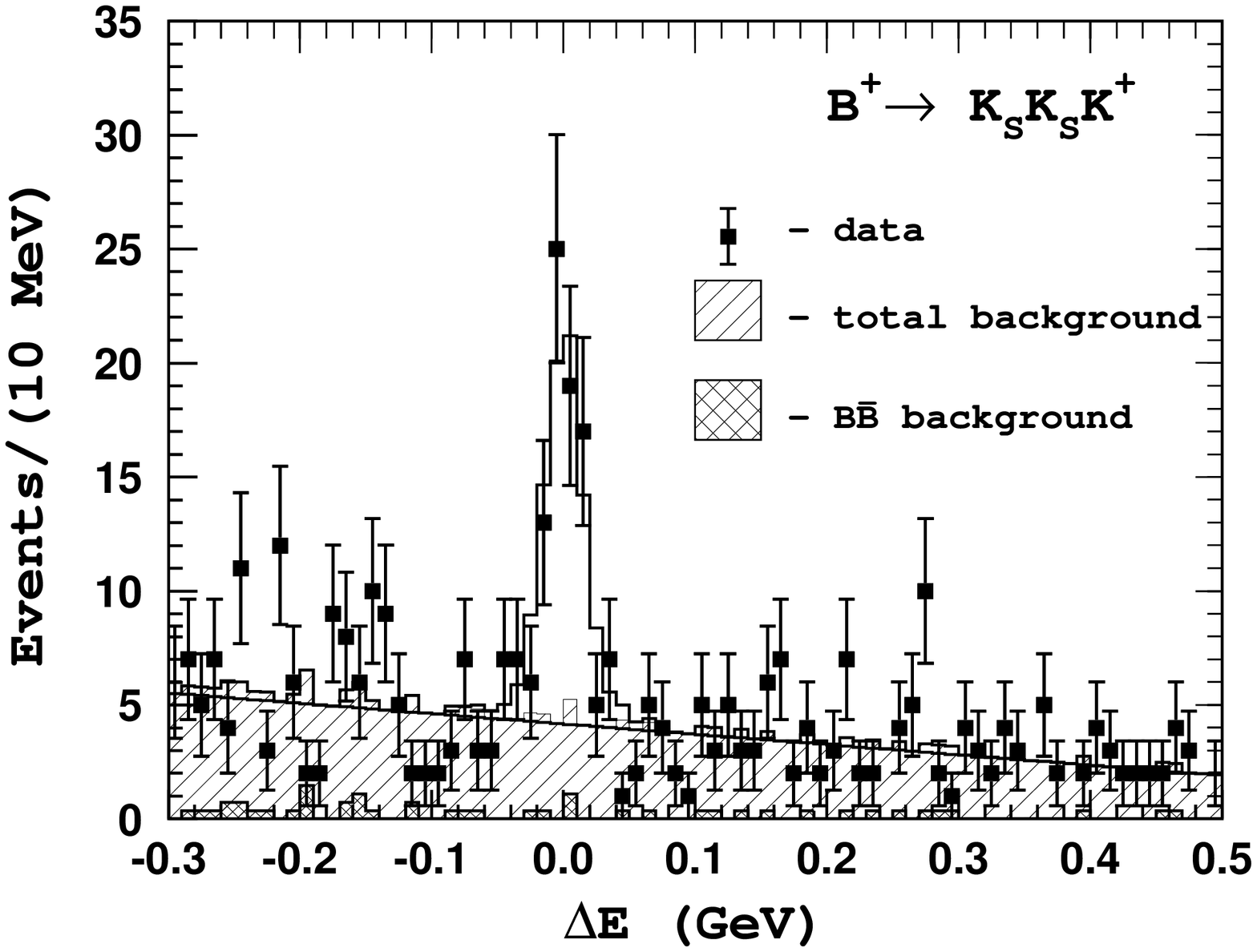} \hfill
  \includegraphics[width=0.49\textwidth]{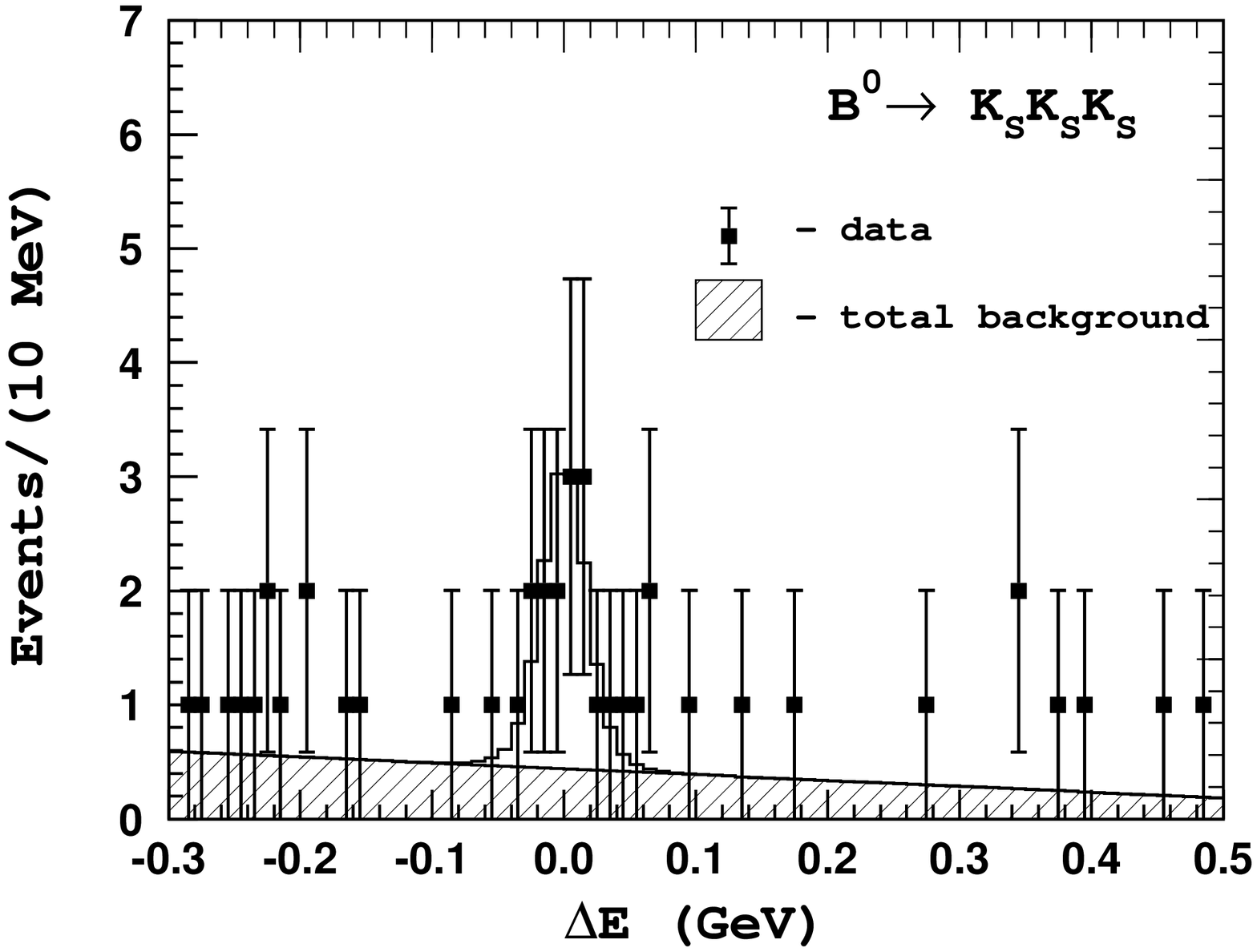}\\
  \caption{$\Delta E$ distributions for $B\to KKK$ three-body final states.
           Points with error bars are data; the open histogram is the fit result;
           the hatched histogram is the background. The straight line shows the
           $q\bar{q}$ continuum background contribution.}
  \label{fig:kkkDE}
\end{figure}

  The $\Delta E$ distributions for all three-kaon final states are shown in
Fig.~\ref{fig:kkkDE}, where data (points with errors) are shown along with the 
background expectation (hatched histograms). To extract the three-body signal 
yields we fit the $\Delta E$ distributions. The results of the fits are 
summarized in Table~\ref{tab:defitall}.
The statistical significance of the $B^0\to K_SK_SK_S$ signal, in terms of the
number of standard deviations is 4.3$\sigma$. It is calculated as 
$\sqrt{-2\ln({\cal{L}}_0/{\cal{L}}_{\rm max})}$, where ${\cal{L}}_{\rm max}$
and ${\cal{L}}_{0}$ denote the maximum likelihood with the nominal signal 
yield and with the signal yield fixed at zero, respectively.
The significance of the signal in all other three-kaon final states exceeds
10$\sigma$. For convenience, some of the results obtained in Ref.~\cite{bc226}
are also given in Table~\ref{tab:defitall}.

\begin{figure}[t]
  \includegraphics[width=80mm,height=63mm]{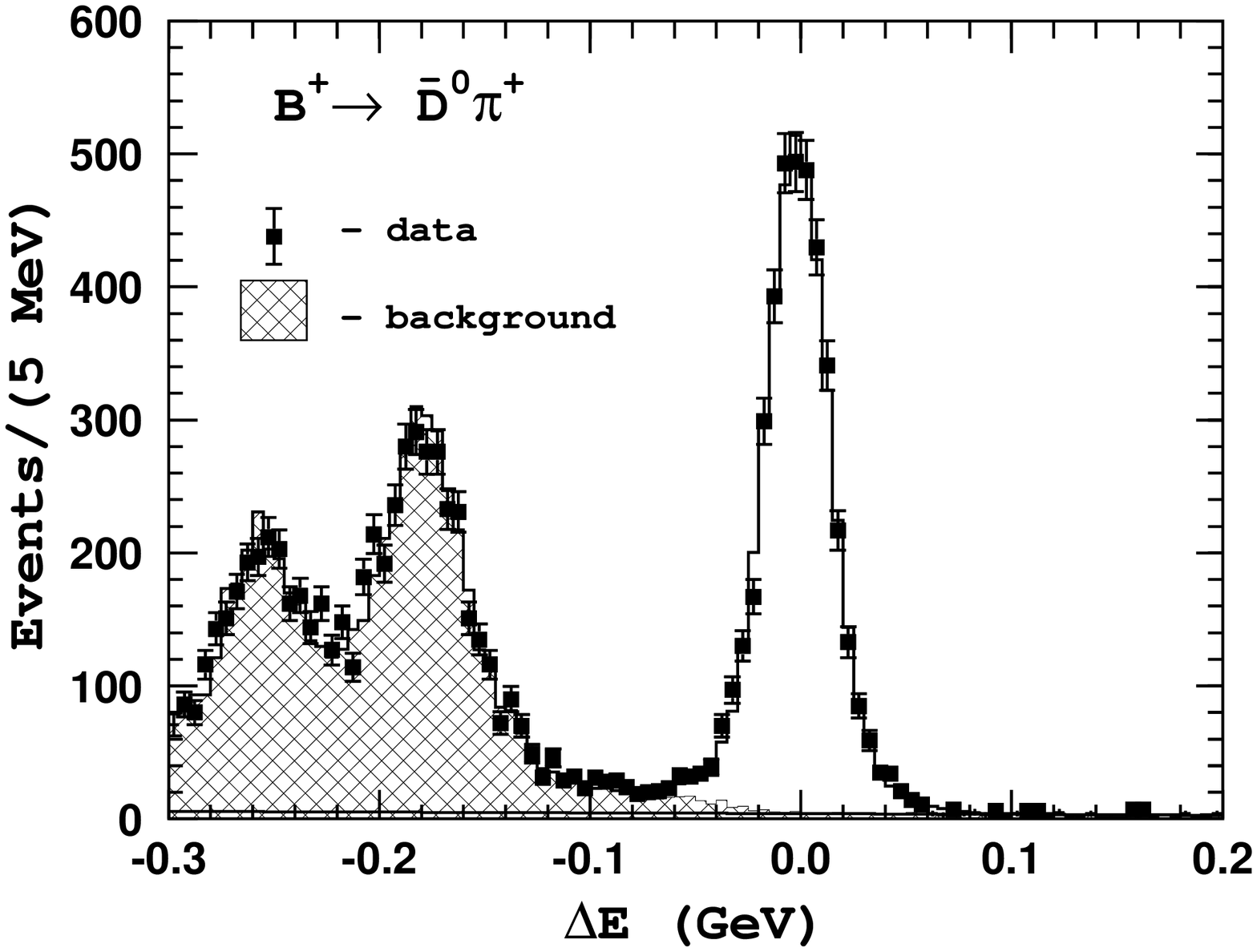} \hfill
  \includegraphics[width=80mm,height=63mm]{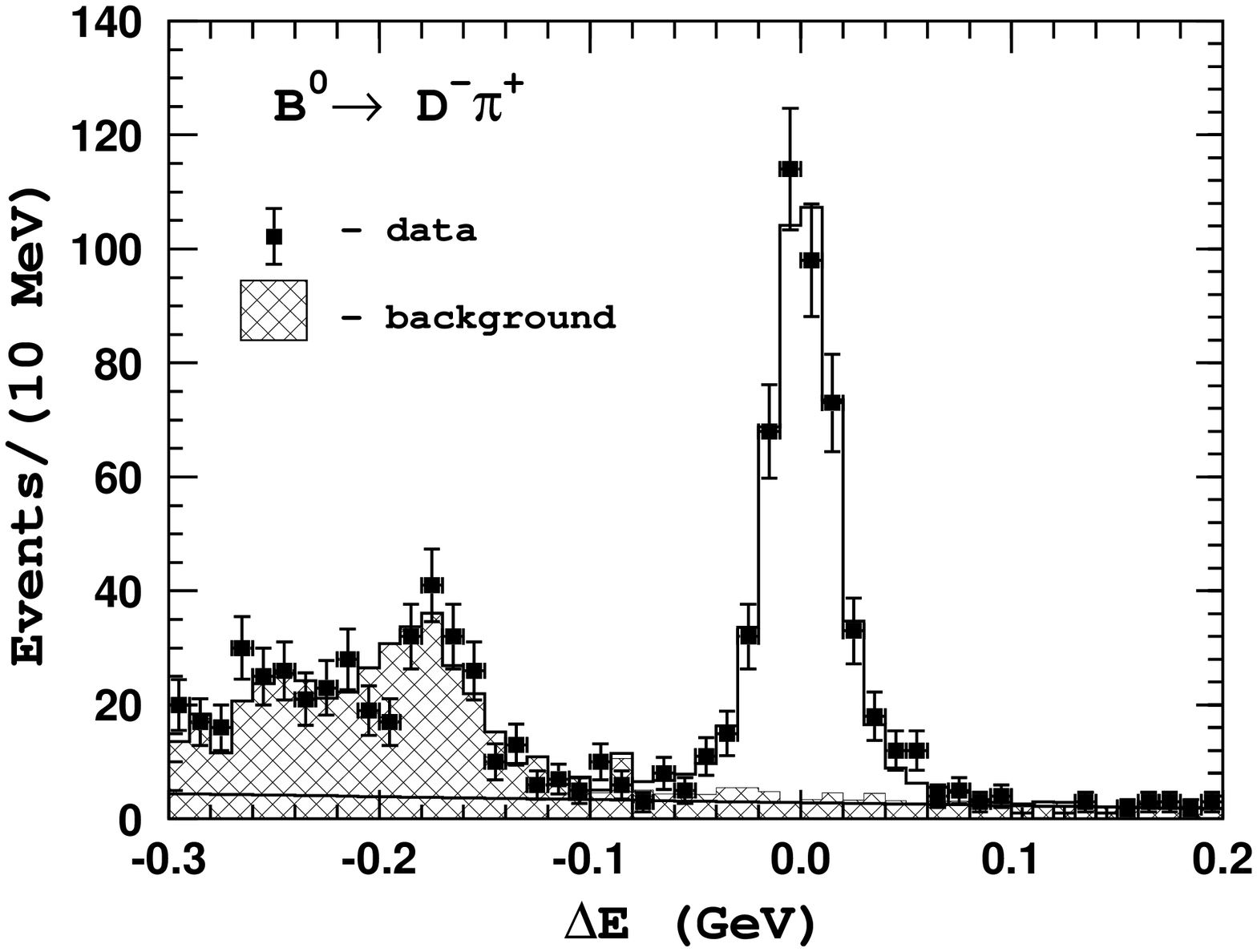}\\
  \caption{$\Delta E$ distributions for $B^+\to\bar{D}^0\pi^+$, 
           $\bar{D}^0\to K^+\pi^-$ (left) and  $B^0\to D^-\pi^+$, 
           $D^-\to K^0\pi^-$ (right) events. The open histogram is the fit result
           and the hatched histogram is the background. The straight line shows 
           the $q\bar{q}$ continuum background contribution.}
  \label{fig:dpiDE}
\end{figure}

   To determine branching fractions, we normalize our results to the observed
$B^+\to\bar{D}^0\pi^+$, $\bar{D}^0\to K^+\pi^-$ and 
$B^0\to D^-\pi^+$, $D^-\to K^0\pi^-$ signals.
 This reduces the systematic errors associated with 
the particle identification efficiency, charged track reconstruction efficiency,
and the event shape variables requirements. We calculate the branching fraction
for $B$ meson decay to a particular final state $f$ via the relation

\[ {\cal{B}}(B\to f) = 
   \frac{N_f}{N_{D\pi}}\frac{\varepsilon_{D\pi}}{\varepsilon_{f}}\times
   {\cal{B}}(B\to D\pi){\cal{B}}(D\to K\pi),
\]
where $N_f$ and $N_{D\pi}$ are the numbers of reconstructed signal events for 
the final state $f$ and that for the $D\pi$ reference process, and 
$\varepsilon_{f}$ and $\varepsilon_{D\pi}$ are the corresponding reconstruction 
efficiencies determined from MC. We use the recently
updated results on $B^+\to \bar{D}^0\pi^+$ and $B^0\to D^-\pi^+$ branching 
fractions from the CLEO Collaboration~\cite{b2dpi_cleo}.
 The $\Delta E$ distributions for the reference
processes $B^+\to\bar{D}^0\pi^+$, $\bar{D}^0\to K^+\pi^-$ and 
$B^0\to D^-\pi^+$, $D^-\to K^0\pi^-$ are shown in Fig.~\ref{fig:dpiDE}.
The results of the fits are summarized in Table~\ref{tab:defitall}.

  The results of the three-kaon branching fraction measurements are presented 
in Table~\ref{tab:defitall}. To determine the reconstruction efficiencies for 
$K^+K^+K^-$ and $K^0K^+K^-$ final states, we use a simple model~\cite{b2khh} 
that takes into account the non-uniform distribution of signal events over the 
Dalitz plot. The three-body signal in this model is parameterized by a $\phi K$
intermediate state and a $f_X K$ state, where $f_X$ is a hypothetical wide 
scalar state. For the $K_SK_SK^+$ and $K_SK_SK_S$ final states, the 
reconstruction efficiencies are determined from MC simulated events that are 
generated with a uniform (phase space) distribution over the Dalitz plot. 
The dominant sources of systematic error are listed in Table~\ref{khh_syst}. 
We estimate the  systematic uncertainty due to variations of reconstruction 
efficiency over the Dalitz plot by varying the relative fractions of 
quasi-two-body states. The uncertainty in the reconstruction efficiencies due
to the limited MC statistics is also included in systematic error.
 The uncertainty due to the particle identification is 
estimated using pure samples of kaons and pions from $D^0\to K^-\pi^+$ decays,
where the $D^0$ flavor is tagged using $D^{*+}\to D^0\pi^+$ decays. The 
systematic error due to uncertainty in the $K_S$ reconstruction efficiency is 
estimated from the study of $D^{*+}\to D^0\pi^+$, $D^0\to K_S\pi^+\pi^-$ decays.
We estimate the uncertainty due to the signal shape parameterization by
varying the parameters of the fitting function within their errors.

  The BaBar Collaboration has recently presented results on charmless three-body
$B$ decays~\cite{babar-hhh}. The reported value, 
$\BF(B^+\to K^+K^+K^-) = (34.7\pm2.0\pm1.8)\times10^{-6}$, is in agreement with
the result presented here and our previously published measurement~\cite{b2khh}.

\begin{table*}[t]              
\caption{List of systematic errors (in percent) for the $B\to KKK$ branching
         fractions.}
\medskip
\label{khh_syst}
  \begin{tabular}{lcccc}  \hline \hline
 Source  &~~$K^+K^+K^-$~&~~$K_SK^+K^+$~&~~$K_SK_SK^+$~&~~$K_SK_SK_S$~ \\ \hline 

 $B\to D\pi$ and $D\to K\pi$ branching fractions & 
                7.7     &      11.0     &      7.7     &    11.0   \\
 Efficiency non-uniformity over the Dalitz plot  & 
                2.2     &       3.6     &       -      &      -    \\
 Signal parameterization                         & 
                2.1     &       5.6     &     4.8      &     7.6   \\
 Particle identification                         & 
                4.0     &       4.0     &      2.0     &     2.0   \\
 $K_S$ reconstruction                            & 
                 -      &        -      &      5.0     &    10.0   \\
 MC statistics                                   &       
                2.1     &       2.4     &      2.8     &     4.3   \\ \hline
    Total   &   9.4     &      13.7     &     10.9     &    17.4   \\
\hline \hline
  \end{tabular}
\end{table*}
%



\begin{figure}[t]
  \centering
  \includegraphics[width=100mm]{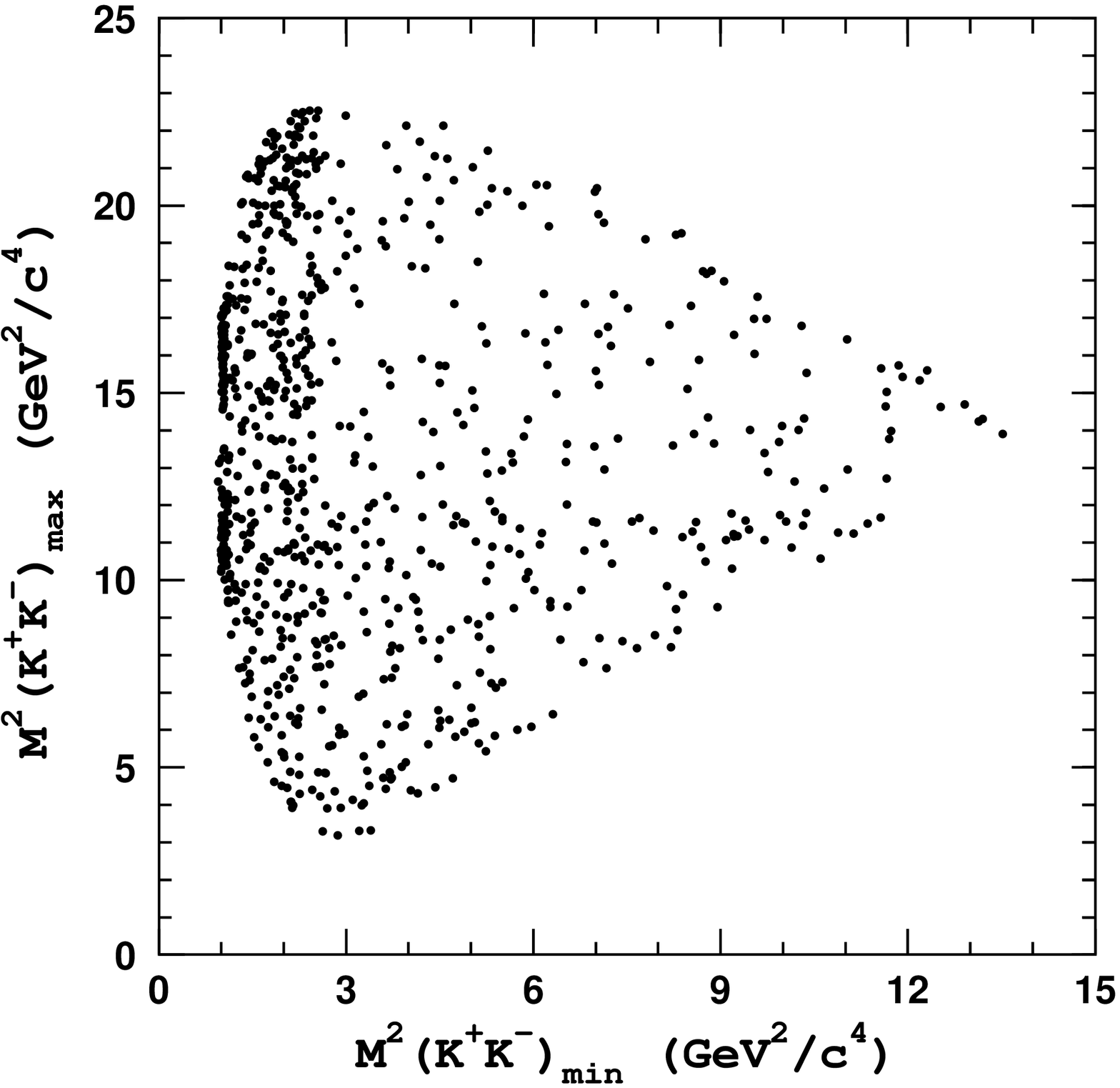}
  \caption{Dalitz plot for $B^+ \to K^+K^+K^-$ candidates in the $B$ 
           signal region.}
  \label{fig:kkk_dp}
  \includegraphics[width=0.49\textwidth]{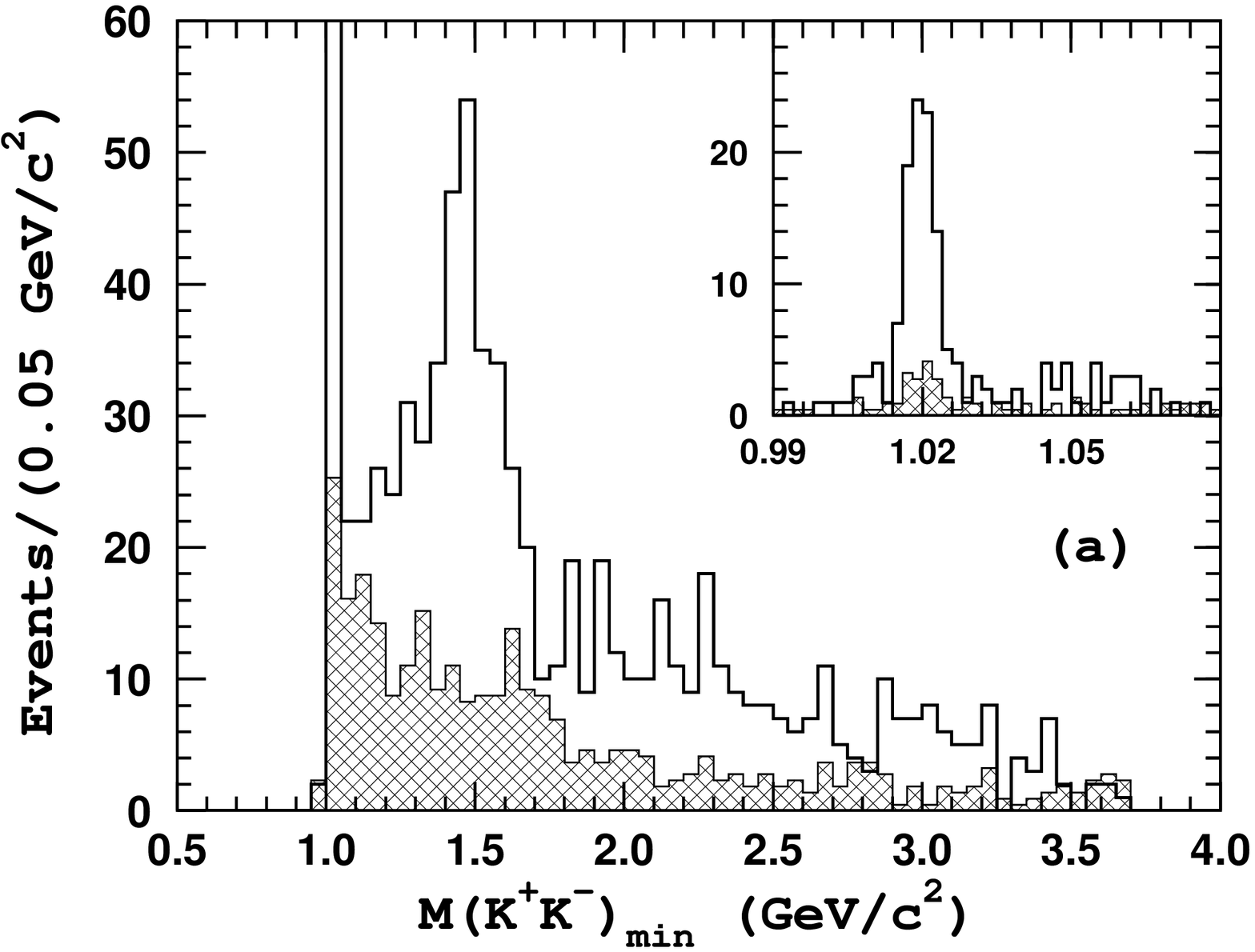} \hfill
  \includegraphics[width=0.49\textwidth]{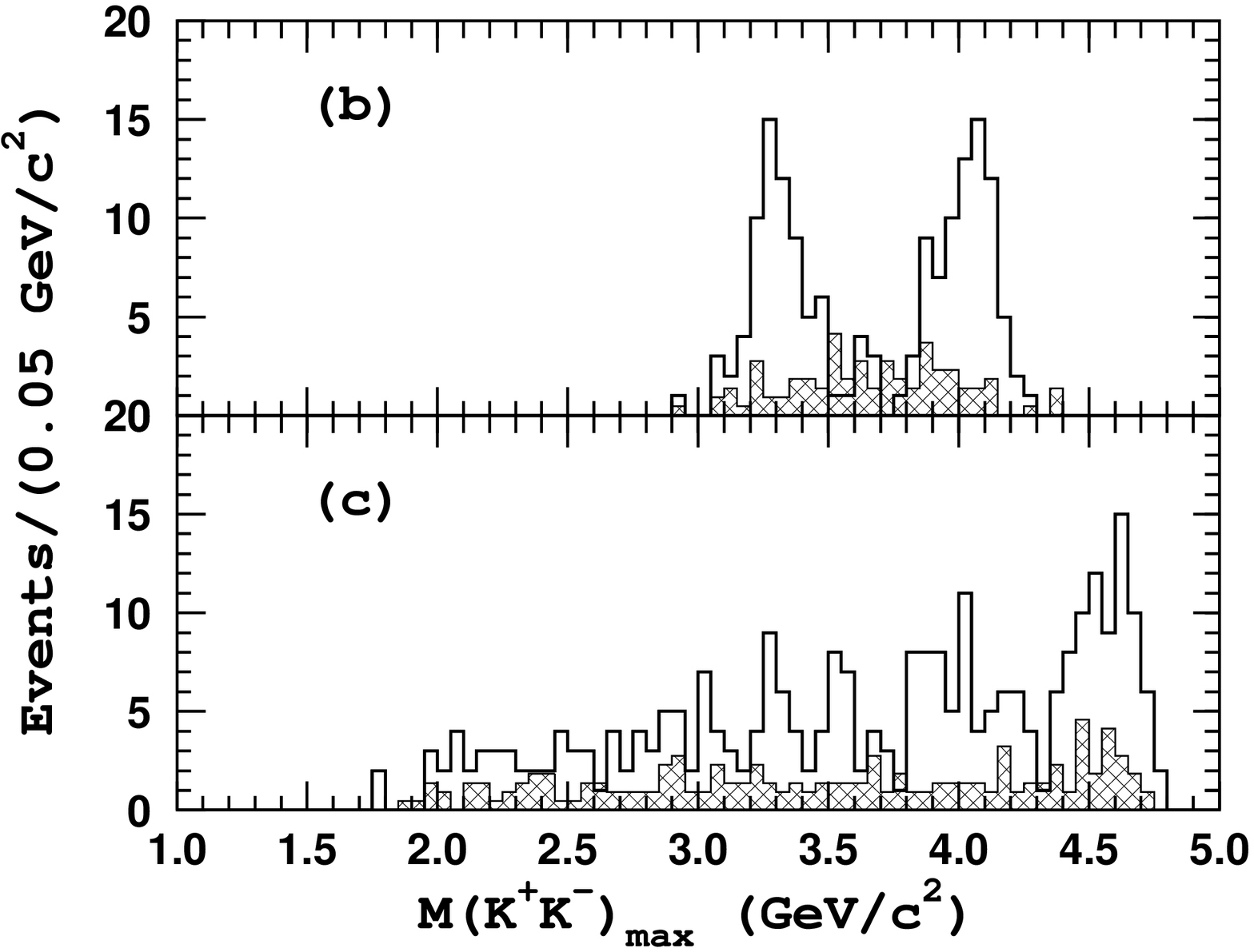}
  \caption{Two-particle invariant mass spectra for $B^+ \to K^+K^+K^-$ 
           candidates from the $B$ signal region (open histograms) and for
           background events in the $\Delta E$ sidebands (hatched histograms). 
           (a) $M(K^+K^-)_{\rm min}$ invariant mass spectrum. The inset in (a)
               shows the $\phi(1020)$ mass region in 2 MeV/$c^2$ bins.
           (b) $M(K^+K^-)_{\rm max}$ spectrum with 
               $M(K^+K^-)_{\rm min}<1.1$~GeV/$c^2$ and 
           (c) $M(K^+K^-)_{\rm max}$ with 
               1.1~GeV/$c^2$ $< M(K^+K^-)_{\rm min}<2.0$~GeV/$c^2$.}
  \label{fig:kkk_kk}
\end{figure}

   To examine possible quasi-two-body intermediate states in the observed 
$B\to KKK$ signals, we analyze the two-kaon invariant mass spectra. The Dalitz
plot for $B^+\to K^+K^+K^-$ candidate events in the $M_{\rm bc}$-$\Delta E$
signal region is shown in Fig.~\ref{fig:kkk_dp}. Since there are two same-charge
kaons in this case, we distinguish the $K^+K^-$ combinations with smaller, 
$M(K^+K^-)_{\rm min}$, and larger, $M(K^+K^-)_{\rm max}$, invariant masses. 
We avoid double entries by forming the Dalitz plot as $M^2(K^+K^-)_{\rm max}$ 
versus $M^2(K^+K^-)_{\rm min}$.

\begin{figure}[t]
  \centering
  \includegraphics[width=100mm]{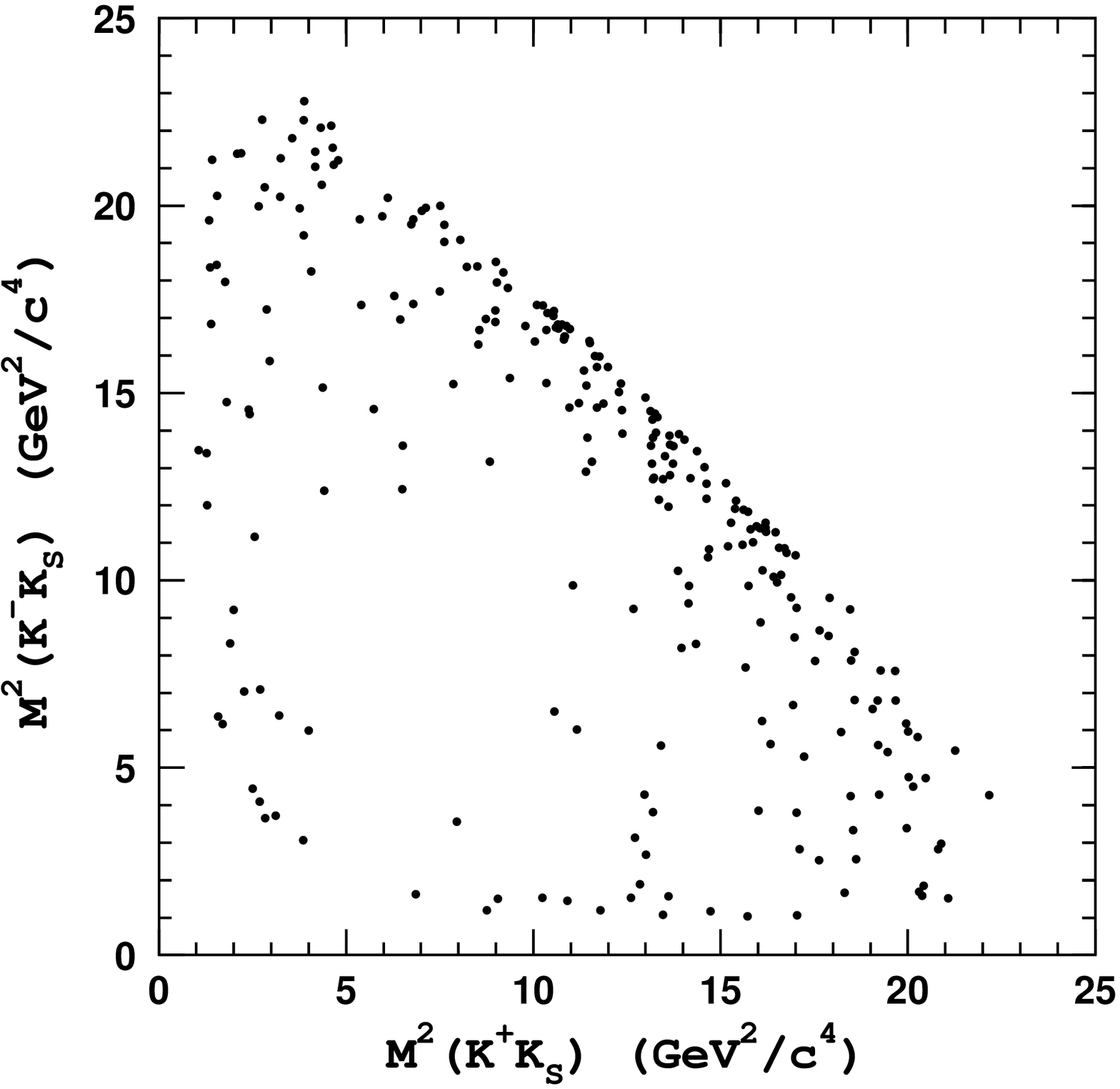}
  \caption{Dalitz plot for $B^0 \to K_SK^+K^-$ candidates in the $B$ 
           signal region.}
  \label{fig:kskk_dp}
  \includegraphics[width=0.49\textwidth]{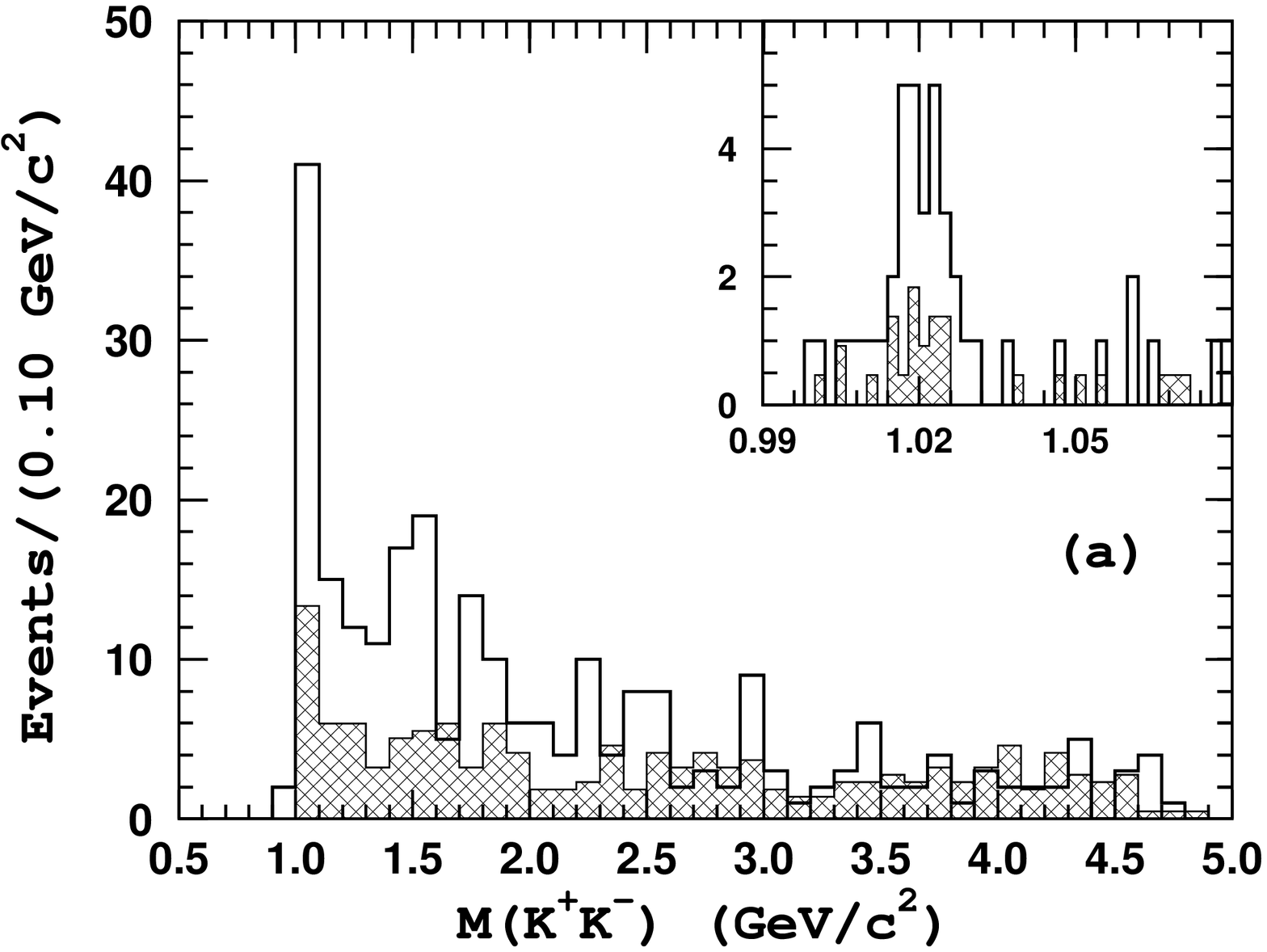} \hfill
  \includegraphics[width=0.49\textwidth]{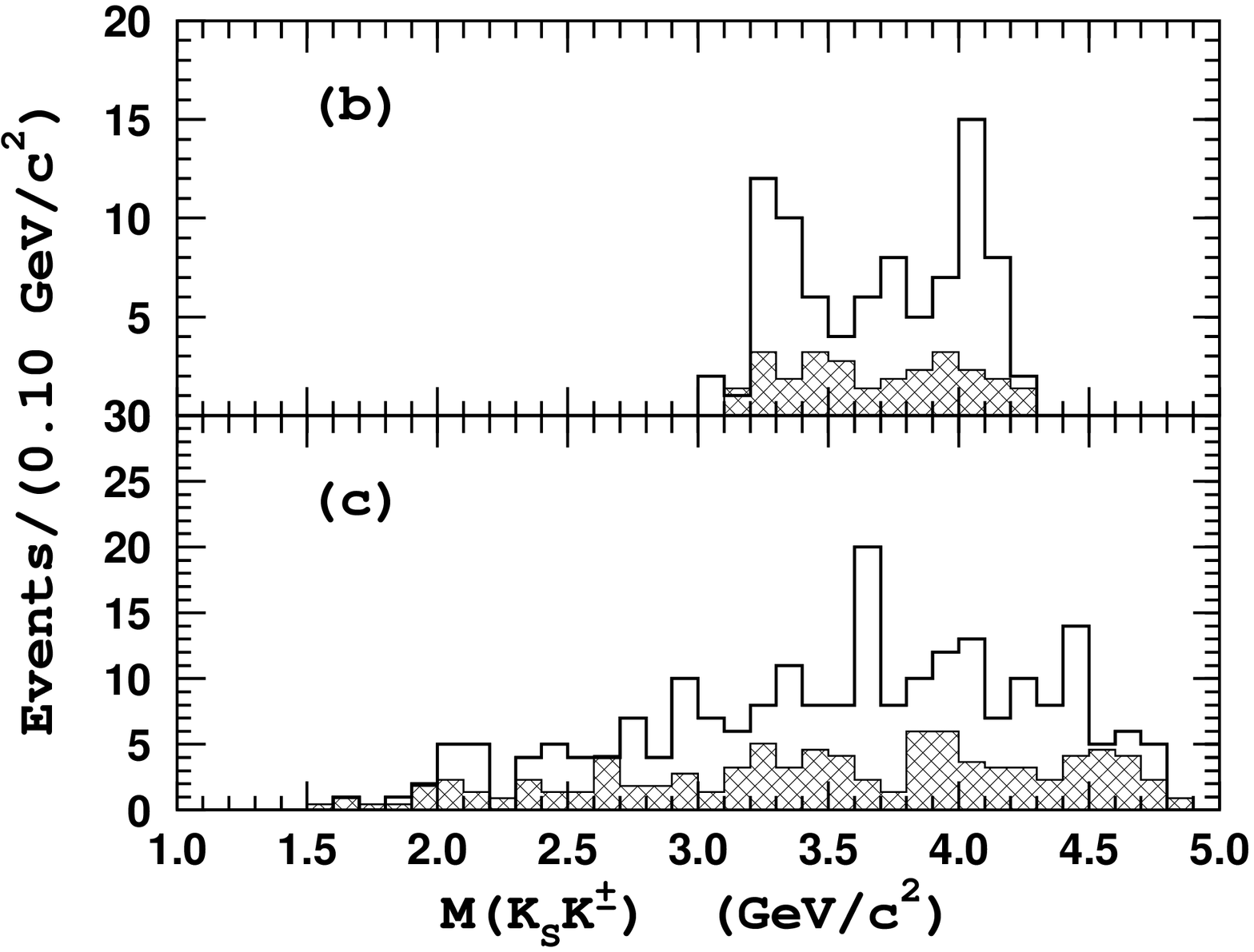} \\
  \caption{Two-particle invariant mass spectra for $B^0 \to K_SK^+K^-$ 
           candidates from the $B$ signal region (open histograms) and for
           background events in the $\Delta E$ sidebands (hatched histograms).
           (a) $K^+K^-$ invariant mass spectrum. The inset in (a) shows the 
               $\phi(1020)$ mass region in 2 MeV/$c^2$ bins.
           (b) $M(K_SK^{\pm})$ spectrum with $M(K^+K^-)<1.1$~GeV/$c^2$ and 
           (c) $M(K_SK^{\pm})$ with
               1.1~GeV/$c^2$ $< M(K^+K^-)_{\rm min}<2.0$~GeV/$c^2$.
           There are two entries per $B$ candidate in (b) and (c).}
  \label{fig:kskk_kk}
\end{figure}
  The $K^+K^-$ invariant mass spectra for events from the $B$ signal region
are shown as open histograms in Figs.~\ref{fig:kkk_kk}(a)-\ref{fig:kkk_kk}(c).
The hatched histograms show the corresponding spectra for background events
in the $\Delta E$ sidebands, normalized to the estimated number of background 
events from the $\Delta E$ fit.
The $M(K^+K^-)_{\rm min}$ spectrum, shown in Fig.~\ref{fig:kkk_kk}(a), is 
characterized by a narrow peak at 1.02~GeV/$c^2$ corresponding to the 
$\phi(1020)$ meson and a broad structure around 1.5~GeV/$c^2$ that is consistent
with a scalar state. In contrast to the $B^+\to K^+\pi^+\pi^-$ three-body 
decay~\cite{bc226}, we also observe a strong
``non-resonant'' enhancement in the $K^+K^+K^-$ final state that extends over
the full $M(K^+K^-)_{\rm min}$ mass range in Fig.~\ref{fig:kkk_kk}(a). To plot
the $M(K^+K^-)_{\rm max}$ mass spectrum we subdivide the $M(K^+K^-)_{\rm min}$
mass region into two ranges: $M(K^+K^-)_{\rm min}<1.1$~GeV/$c^2$ and 
1.1~GeV/$c^2$ $< M(K^+K^-)_{\rm min}<2.0$~GeV/$c^2$. The $M(K^+K^-)_{\rm max}$
mass spectra for these two regions are shown separately in 
Fig.~\ref{fig:kkk_kk}(b) and Fig.~\ref{fig:kkk_kk}(c), respectively. The 
prominent two-peak structure apparent in Fig.~\ref{fig:kkk_kk}(b) is due to 
the 100\% $\phi$ meson polarization in the $B^+\to\phi K^+$ decay, which 
results from angular momentum conservation.

  The Dalitz plot for $B^0\to K_SK^+K^-$ candidate events in the 
$M_{\rm bc}$-$\Delta E$ signal region is shown in Fig.~\ref{fig:kskk_dp}. 
The $KK$ invariant mass spectra for these events are shown as open histograms
in Fig.~\ref{fig:kskk_kk}, and the hatched histograms show the corresponding
spectra for background events in the $\Delta E$ sidebands. The structure 
observed in the $K^+K^-$ mass spectrum is very similar to that observed in 
$M(K^+K^-)_{\rm min}$ spectrum for the $K^+K^+K^-$ final state 
(see Fig.~\ref{fig:kkk_kk}): a prominent peak that corresponds to the $\phi$ 
meson and a broad enhancement in the higher $K^+K^-$ mass region.

\begin{figure}[t]
  \centering
  \includegraphics[width=0.49\textwidth]{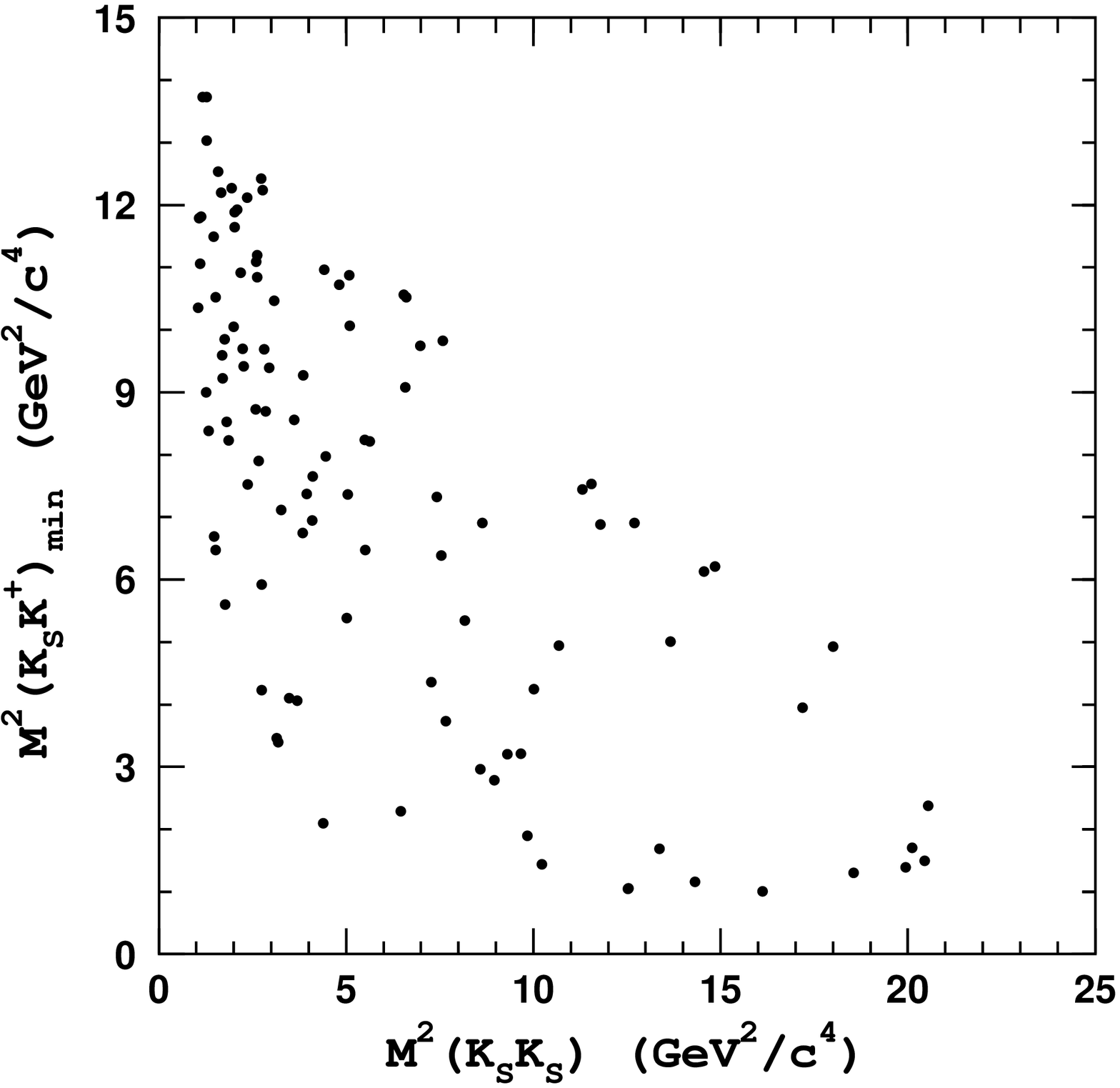}\hfill
  \includegraphics[width=0.49\textwidth]{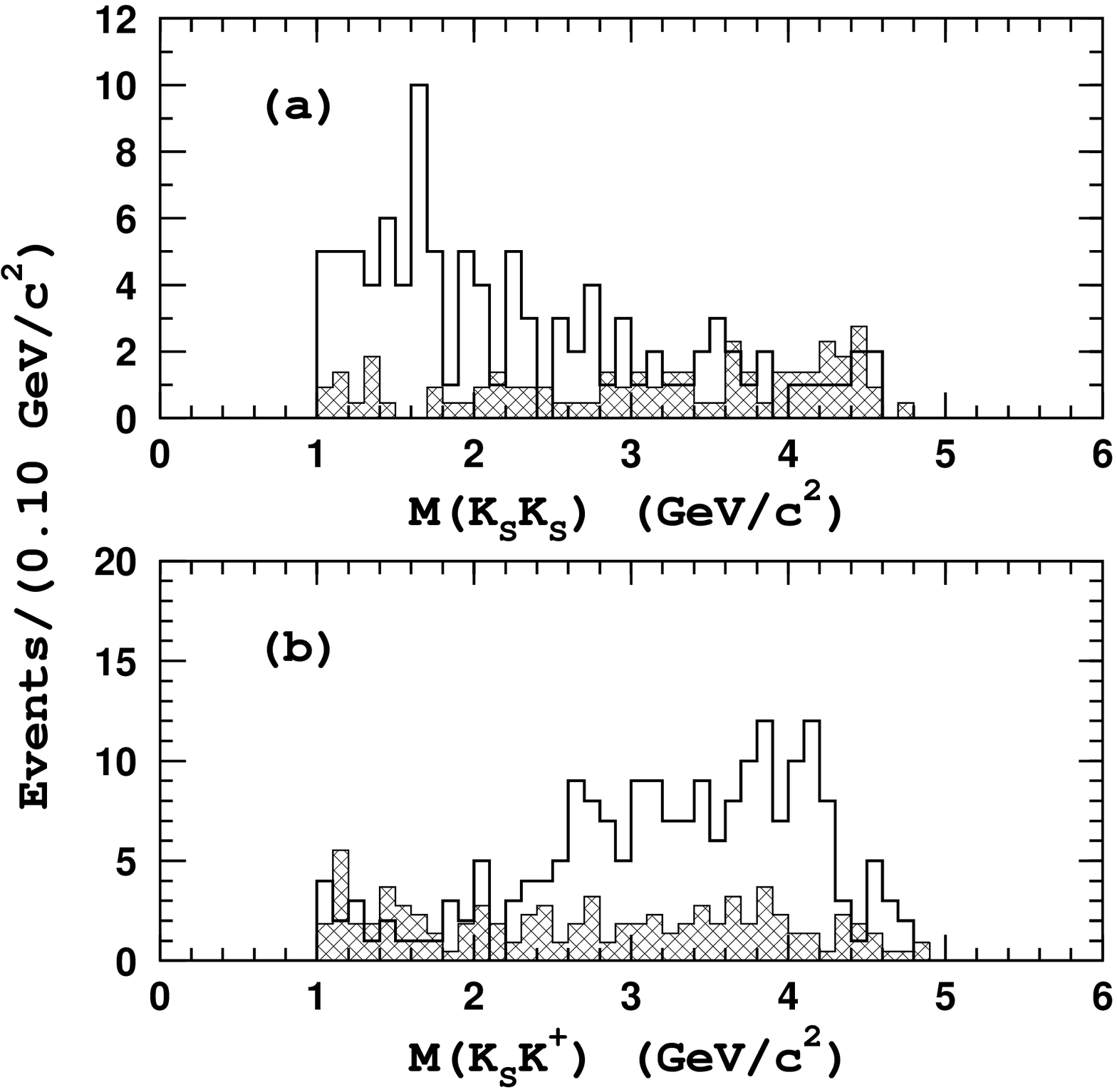}\\
  \caption{Dalitz plot and two-particle invariant mass spectra for 
           $B^+ \to K_SK_SK^+$ candidates in the $B$ signal region. 
           There are two entries per $B$ candidate in (b).}
  \label{fig:ksksk}
\end{figure}

  The numbers of reconstructed $B^+\to K_SK_SK^+$ and $B^0\to K_SK_SK_S$ signal
events are significantly smaller because of the additional suppression due to
the $K^0\to K_S \to \pi^+\pi^-$ branching fraction. A Dalitz plot for
the $B^+\to K_SK_SK^+$ candidate events in the $M_{\rm bc}$-$\Delta E$ signal
region is shown in Fig.~\ref{fig:ksksk}, along with two-kaon invariant mass 
spectra. It is apparent in Fig.~\ref{fig:ksksk} that
the $K_SK_S$ invariant mass spectrum is quite similar to the $K^+K^-$ mass 
spectrum observed in the $K_SK^+K^-$ final state. Except for the absence of 
the $\phi$ meson, which cannot decay to $K_SK_S$, we observe a similar broad
structure in the higher $K_SK_S$ mass region. The low mass region of the 
$K_SK^+$ mass spectrum shown in Fig.~\ref{fig:ksksk}(b) agrees with background
and exhibits no prominent structures.

\section{Discussion \& Conclusion}

  Charmless $B$ meson decays have attracted a considerable amount of attention
in recent years, primarily because they provide a possible way to extract weak
phases. An important check of the Standard Model would be provided by
measurements of the same CP-violating parameter in different weak interaction 
processes. 
A good example is the comparison of the measurement of the coefficient of the
CP violating $\sin(\Delta m_dt)$ term in the time dependent analysis of neutral
$B$ meson decays. In $B^0\to (c\bar{c})K^0$ decays (where $(c\bar{c})$ denotes
a charmonium state) this coefficient is $\sin(2\phi_1)$. Precise measurements 
of $\sin(2\phi_1)$ (also known as $\sin(2\beta)$) have recently been 
reported by the Belle and BaBar experiments~\cite{CP_phi1}. The best known 
candidates for $b\to s$ penguin dominated processes where this quantity can be
measured independently are $B^0\to\phi K^0$ and $B^0\to\eta' K^0$ decays. 
However, these modes have small branching fractions of order $10^{-6}-10^{-5}$
(including secondary branching fractions). Thus, very large numbers of $B$ 
mesons are required to perform these measurements. 
This is especially true for the $\phi K^0$ final state. The large signal 
observed in the three-body $B^0\to K_SK^+K^-$ decay mode, where the $\phi K_S$
two-body intermediate state gives a relatively small contribution, would
significantly increase statistics if these events could be used.
There are two possible complications:
(1) Whilst the $\phi K_S$ state has fixed CP, the CP-parity of the three-body 
$K_SK^+K^-$ final state is not fixed. If the fractions of CP-even and CP-odd 
components are comparable, the $K_SK^+K^-$ state will not be useful for a
CP violation measurement; (2) Possible $b\to u$ tree contributions may introduce 
an additional weak phase in the $B^0\to K_SK^+K^-$ amplitude and cloud the
interpretation of any observed CP violation. The $b\to u$ contribution in 
$B^0\to\phi K_S$ is expected to be negligible (since $\phi$ is a pure $s\bar{s}$
state), but this is not necessarily the case for the three-body $K_SK^+K^-$ 
final state. Here we discuss the possibility of the use of the three-body
$B^0\to K_SK^+K^-$ decay mode for CP measurements.

  The decays of $B$ mesons to three-body $Khh$ final states can be described 
by $b\to u$ tree-level spectator and  $b\to s(d)g$ one-loop penguin 
diagrams. Although $b\to u$ $W$-exchange and annihilation diagrams can also 
contribute to these final states, they are expected to be much smaller and we 
neglect them in the following discussion.

  $B$ meson decays to final states with odd numbers of kaons ($s$-quarks) are
expected to proceed dominantly via the $b\to sg$ penguin transition since, for
these cases, the $b\to u$ tree contribution  has an additional CKM suppression.
In contrast, decays with two final state kaons proceed via the $b\to u$ tree 
transition with no $b\to sg$ penguin contribution. This allows us to estimate
the $b\to u$ tree contribution to final states with three kaons via the analysis
of $KK\pi$ final states. It is also important to note that the $b\to sg$ penguin
transition is an isospin conserving process, while the $b\to u$ tree and 
$b\to dg$ penguin transitions are isospin violating.

\begin{figure}[t]
  \begin{minipage}[c]{1.1\textwidth}
  \hspace*{-1.1cm}\includegraphics[width=0.33\textwidth]{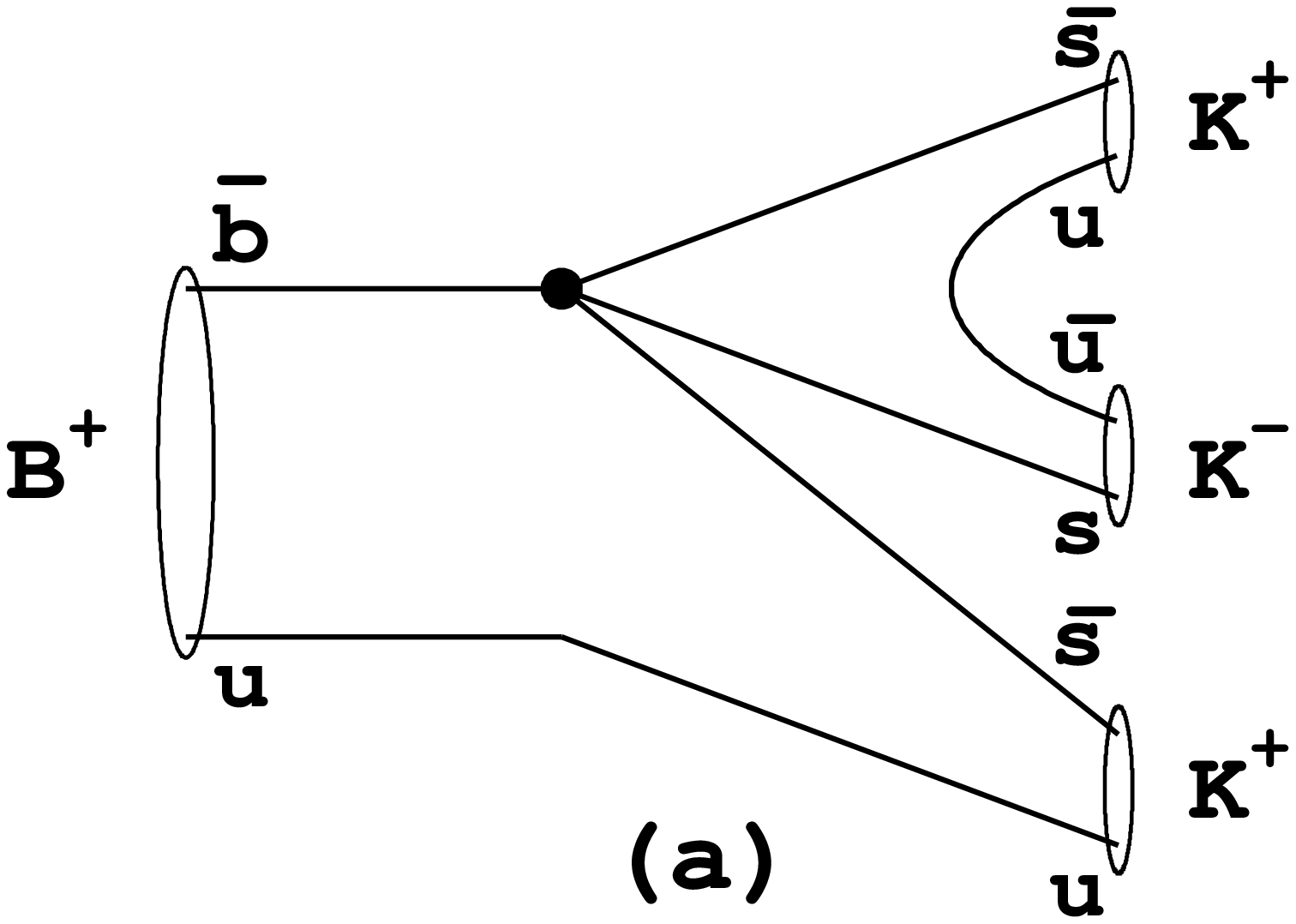}
  \hspace*{-0.3cm}\includegraphics[width=0.33\textwidth]{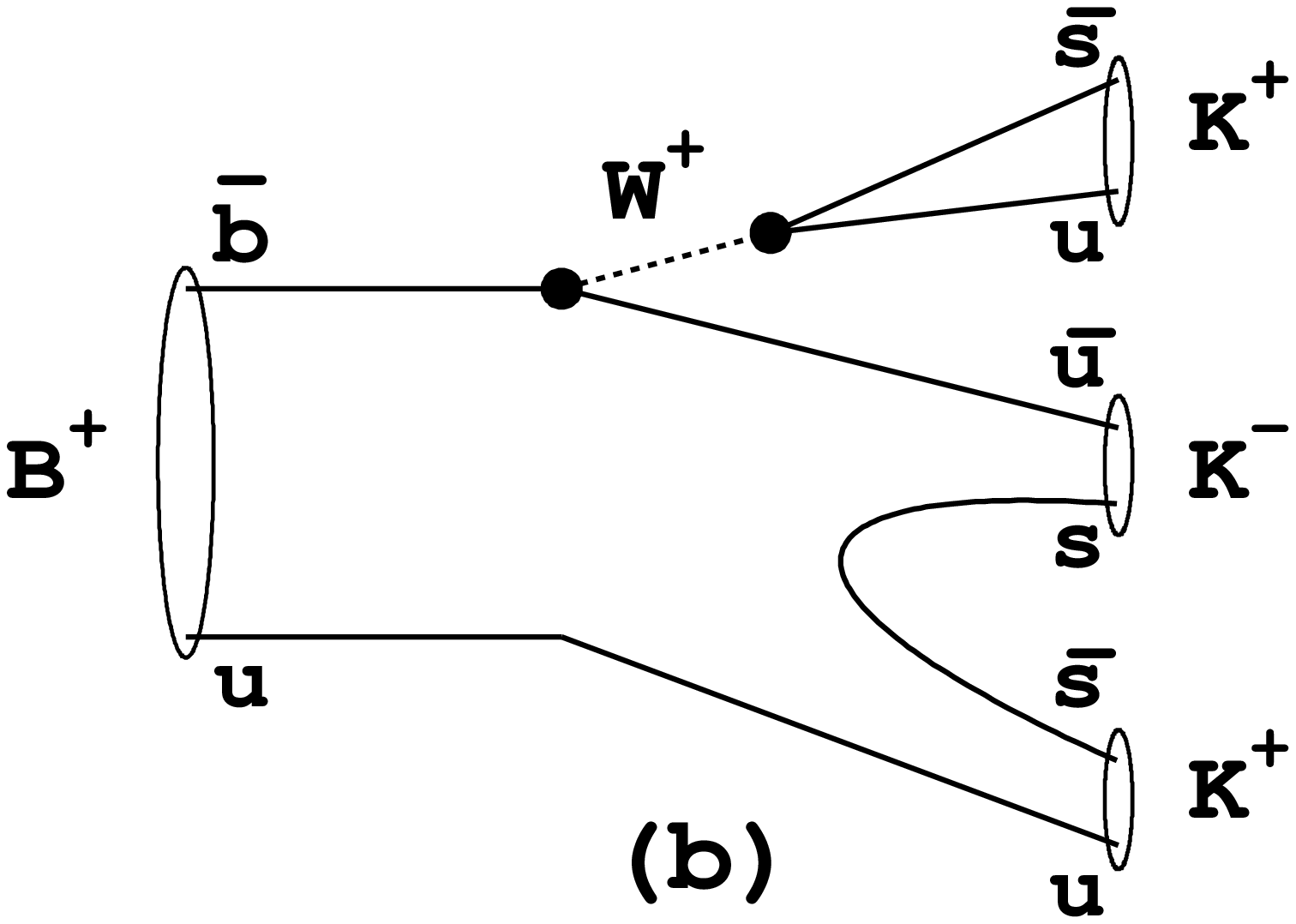}
  \hspace*{-0.6cm}\includegraphics[width=0.33\textwidth]{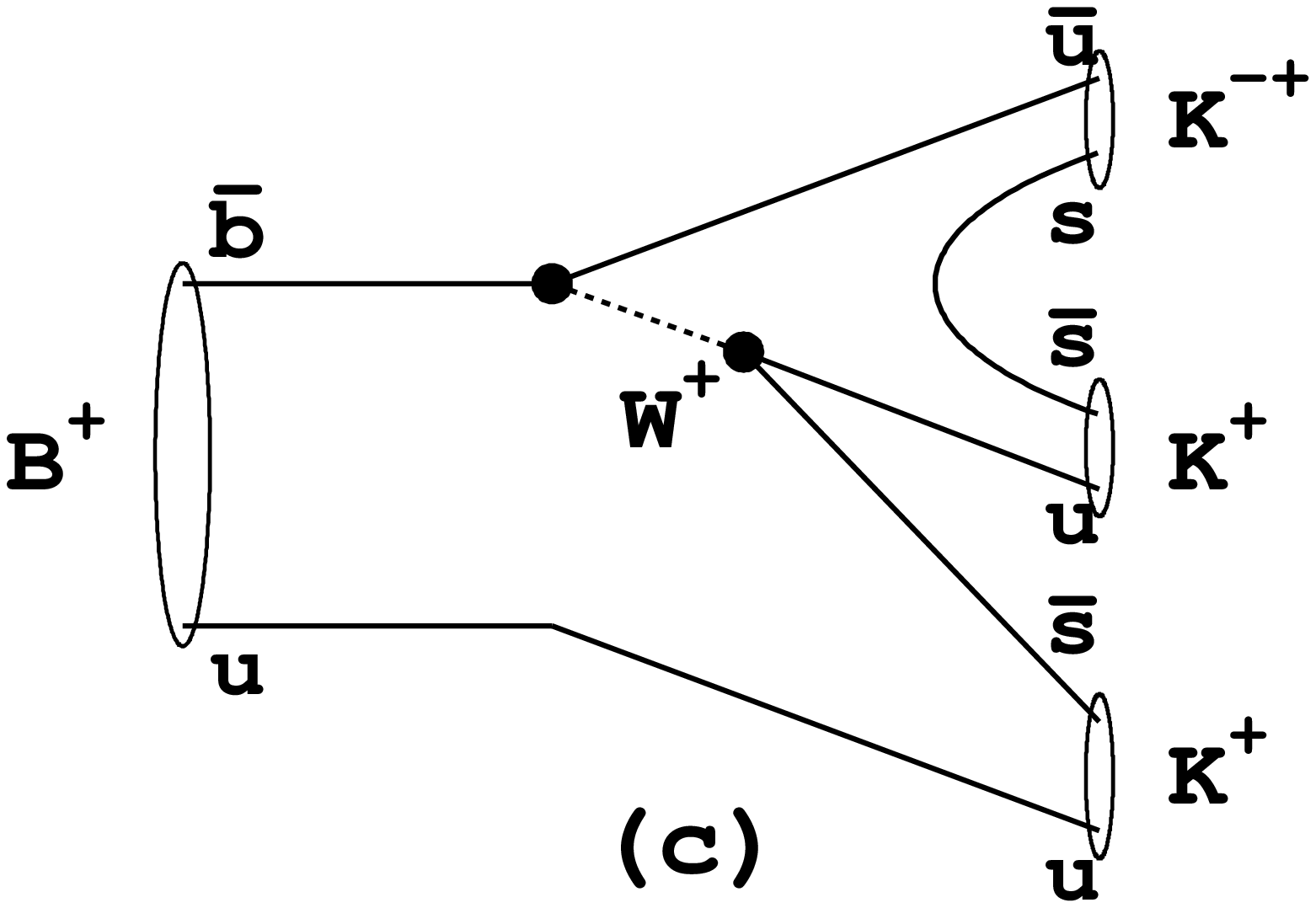}\\
  \hspace*{-1.1cm}\includegraphics[width=0.33\textwidth]{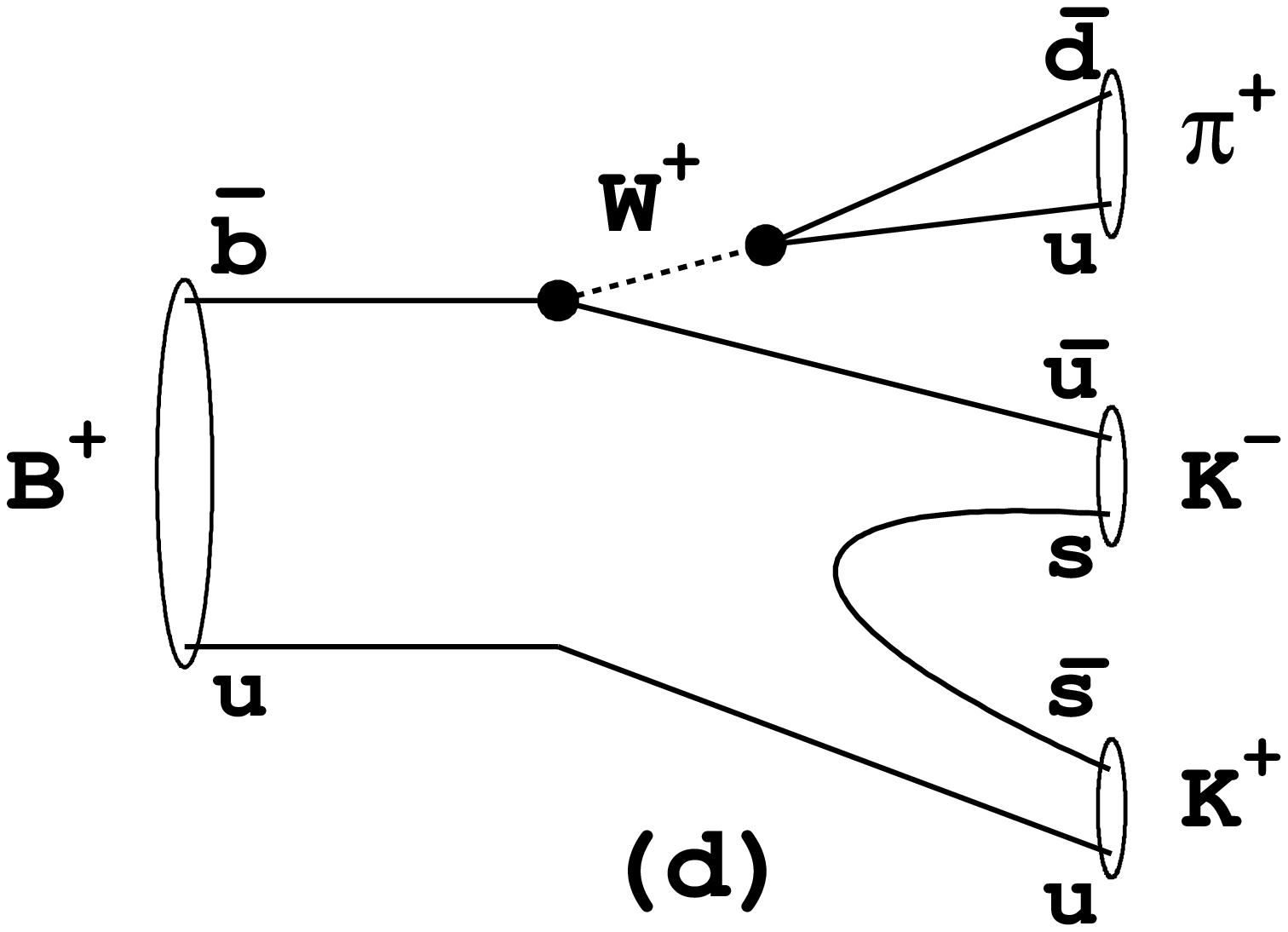}
  \hspace*{-0.3cm}\includegraphics[width=0.33\textwidth]{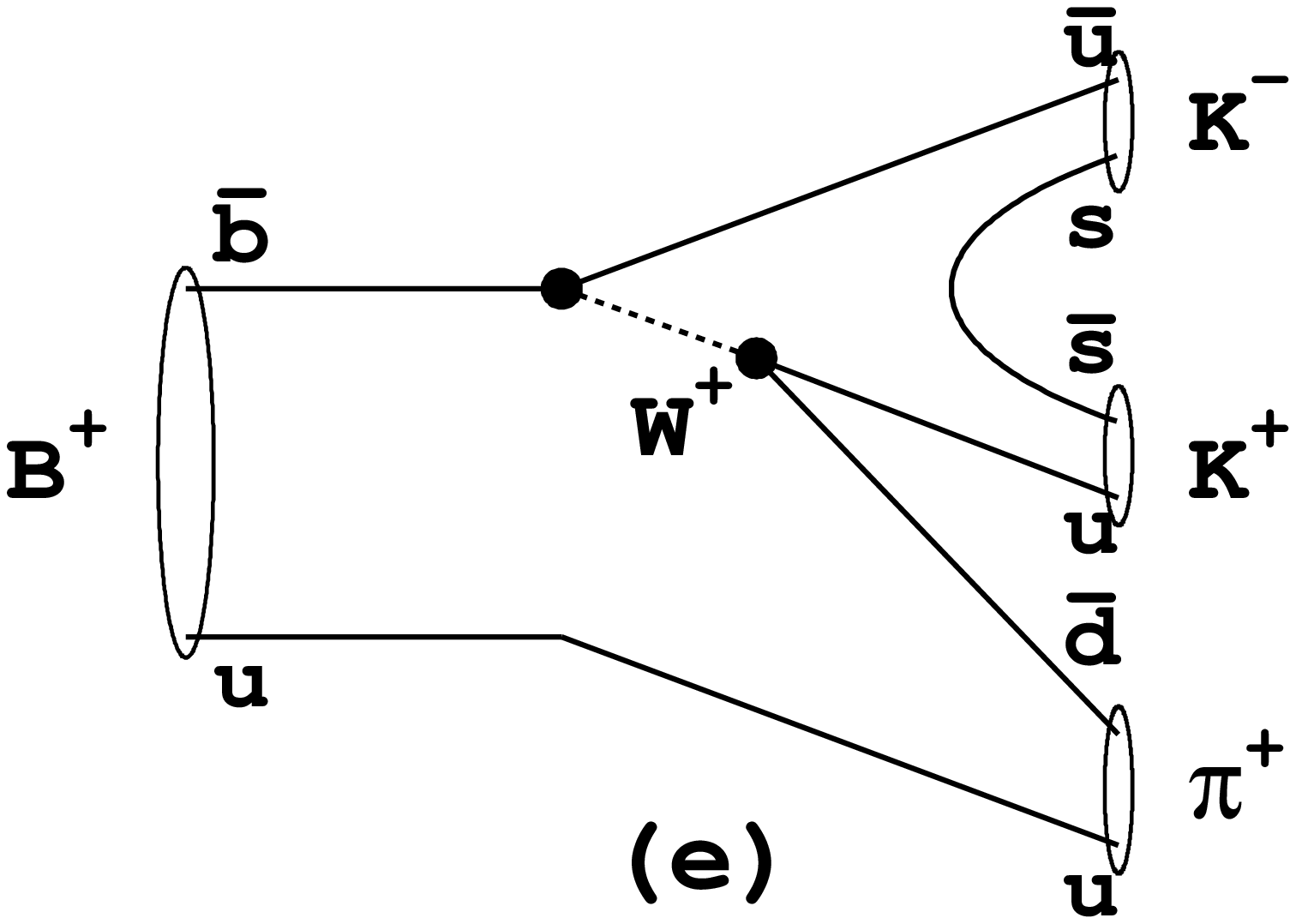}
  \hspace*{-0.6cm}\includegraphics[width=0.33\textwidth]{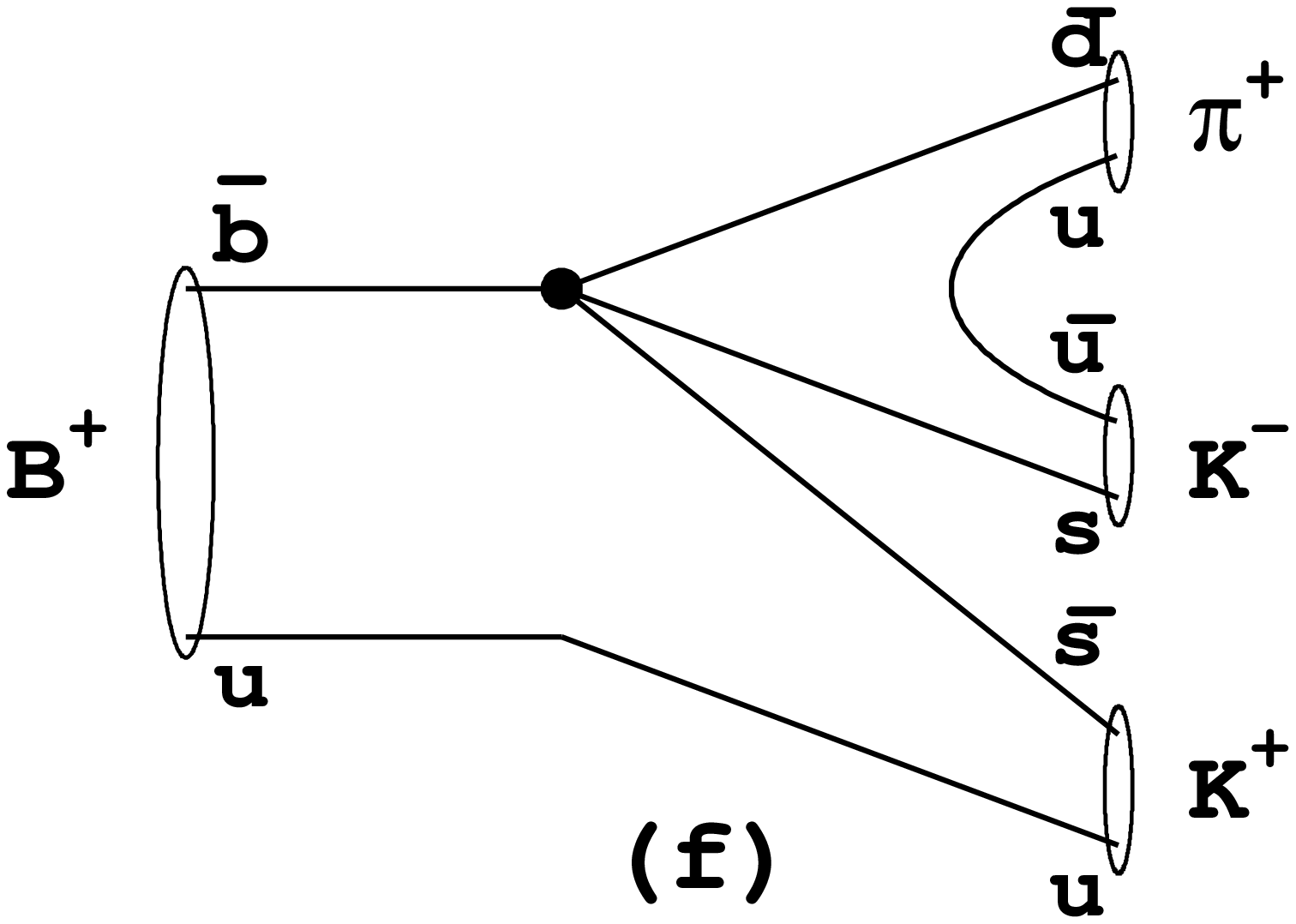}\\
  \end{minipage}
  \begin{minipage}[c]{1.0\textwidth}
  \caption{Diagrams for $B^+\to K^+K^+K^-$ decay: 
           (a) $b\to s$ penguin; (b) and (c) $b\to u$ trees, and 
           for $B^+\to K^+K^-\pi^+$ decay: (d) and (e) $b\to u$ trees;
           (f) $b\to d$ penguin.}
  \label{fig:diagrams}
  \end{minipage}
\end{figure}

  This is illustrated for $B^+\to K^+K^+K^-$ decay in Fig.~\ref{fig:diagrams}.
The diagram that corresponds to the main $b\to s$ penguin contribution is shown
in Fig.~\ref{fig:diagrams}(a). The $b\to u$ tree contribution,
Fig.~\ref{fig:diagrams}(b), has an additional Cabibbo suppression due to the
$W^+\to \bar{s}u$ vertex. The corresponding diagram without Cabibbo 
suppression ($W^+\to \bar{d}u$) is shown in Fig.~\ref{fig:diagrams}(d) and 
expected to be the dominant contributor to the $K^+K^-\pi^+$ final state.
A quantitative estimate of the size of the $b\to u$ tree amplitude is provided
by the ratio

\[
  F \equiv
  \frac{|{\cal{A}}^{KKK}_{b\to u}|^2}{|{\cal{A}}^{KKK}_{\rm total}|^2}\sim
  \frac{{\cal{B}}(B^+\to K^+K^-\pi^+)}{{\cal{B}}(B^+\to K^+K^+K^-)}\times
  \left ( \frac{f_{K}}{f_{\pi}} \right )^2\times\tan^2\theta_C,
\]
where ${\cal{A}}^{KKK}_{\rm total}$ is the total amplitude for the
$B^+\to K^+K^+K^-$ decay and ${\cal{A}}^{KKK}_{b\to u}$ is its $b\to u$ tree 
contribution. The $(f_{K}/f_{\pi})^2$ factor, where $f_{\pi} = 131$~MeV and
$f_K = 160$~MeV are pion and kaon decay constants, respectively, takes into 
account the corrections for SU(3) breaking effects in the factorization 
approximation, and $\theta_C$ is the Cabibbo angle 
($\sin\theta_C=0.2205\pm0.0018$)~\cite{PDG}. Using the results for 
$B^+\to K^+K^-\pi^+$ and $B^+\to K^+K^+K^-$ branching fractions from 
Table~\ref{tab:defitall}, we obtain $F = 0.022\pm0.005$. Similarly, for $B^0$
decays to $K_SK^+K^-$ and $K_SK^+\pi^-$ final states, respectively, we find
$F = 0.023\pm0.013~(<0.037)$, where the second number is obtained using 
the upper limit for the $B^0\to K_SK^+\pi^-$ branching fraction. The small value 
of $F$ indicates that we can
neglect the $b\to u$ tree contribution to the $B\to KKK$ rates and perform an
isospin analysis of the three-kaon final states. From an isospin decomposition
of $B$ mesons wave functions we obtain the following relations between 
three-kaon branching fractions,
\begin{equation}
  {\cal{B}}(B^0\to K^0K^+K^-) = {\cal{B}}(B^+\to K^+K^+K^-)\times
  \frac{\tau_{B^0}}{\tau_{B^+}};
  \label{eq:rel_1}
\end{equation}
\begin{equation}
  {\cal{B}}(B^0\to K^0K^+K^-) = {\cal{B}}(B^+\to K^+K^0\bar{K}^0)\times
  \frac{\tau_{B^0}}{\tau_{B^+}},
  \label{eq:rel_2}
\end{equation}
where the factor $\tau_{B^0}/\tau_{B^+}$ takes into account the difference in
total widths of charged and neutral $B$ mesons. We use the first relation, 
Eq.~\ref{eq:rel_1}, as a check of our assumption that the isospin violating 
contribution is small, and calculate the ratio
\[
  R \equiv 
  \frac{{\cal{B}}(B^0\to K^0K^+K^-)}{{\cal{B}}(B^+\to K^+K^+K^-)}\times
  \frac{\tau_{B^+}}{\tau_{B^0}} = 
  \frac{N_{K_SK^+K^-}}{N_{K^+K^+K^-}}\times
  \frac{\varepsilon_{K^+K^+K^-}}{\varepsilon_{K^0K^+K^-}}\times
  \frac{\tau_{B^+}}{\tau_{B^0}} = 0.95 \pm 0.11 \pm 0.06,
\]
where we use signal yields ($N$) and reconstruction efficiencies 
($\varepsilon$) from Table~\ref{tab:defitall} instead of branching fractions
to reduce the systematic error, and 
$\tau_{B^+}/\tau_{B^0} = 1.091\pm0.023\pm0.014$~\cite{Blife}.
The calculated value agrees with unity within its statistical error.

  Let us now consider the second isospin relation, Eq.~\ref{eq:rel_2}, in more
detail. The $B^+\to K^+K^0\bar{K}^0$
decay results in three different observable states: $K^+K_SK_S$, $K^+K_LK_L$
and $K^+K_SK_L$. The relative fractions of these states depends on the relative
fractions of states with even and odd orbital momenta in the $K^0\bar{K}^0$
system. Bose statistics requires that the $K^0\bar{K}^0$ wave function be 
symmetric (and, therefore, CP even), independently of the relative orbital 
momentum, $l$, of the neutral kaons. As a result, a $K^0\bar{K}^0$ system 
with even orbital momenta can only decay to $K_SK_S$ or $K_LK_L$ final states
(with equal fractions), while a $K^0\bar{K}^0$ system with odd orbital momenta
can only decay to the $K_SK_L$ final state. Thus, the $K^+K^0\bar{K}^0$ wave 
function can be written in the following form

\begin{equation}
  |K^+K^0\bar{K}^0> = \alpha \frac{|K^+K_SK_S> + |K^+K_LK_L>}{\sqrt{2}}+
  \beta |K^+K_SK_L>,
  \label{eq:cp_dec}
\end{equation}
where $\alpha$ and $\beta$ are unknown coefficients constrained by
$\alpha^2+\beta^2=1$.

  In this experiment, we observe only the $K^+K_SK_S$ component of the 
$K^+K^0\bar{K}^0$ final state. Measuring the $B^+\to K^+K^+K^-$ and 
$B^+\to K^+K_SK_S$ branching fractions and using the isospin relation, 
Eq.~\ref{eq:rel_2}, with the wave function decomposition, Eq.~\ref{eq:cp_dec},
we can determine the parameter $\alpha^2$,
\begin{equation}
  \alpha^2 = 
  2\frac{{\cal{B}}(B^+\to K^+K_SK_S)}{{\cal{B}}(B^0\to K^0K^+K^-)}\times
  \frac{\tau_{B^0}}{\tau_{B^+}} = 
  2\frac{N_{K^+K_SK_S}}{N_{K_SK^+K^-}}\times
  \frac{\varepsilon_{K^0K^+K^-}}{\varepsilon_{K^+K_SK_S}}\times
  \frac{\tau_{B^0}}{\tau_{B^+}}.
  \label{eq:alpha_1}
\end{equation}
Here, the parameter $\alpha^2$ characterizes the fraction of states with even
orbital momenta in the $K^0\bar{K^0}$ system in the three-body $K^+K^0\bar{K^0}$
final state. At the same time, due to isospin symmetry, $\alpha^2$ also
gives the fraction of states with even orbital momenta in the $K^+K^-$ system of
the three-body $K^0K^+K^-$ final state. Since the total angular momentum of the
$K^0K^+K^-$ system is zero, the orbital momentum of the $K^+K^-$ pair relative
to the remaining neutral kaon, $l'$, is equal to $l$. Thus, the CP-parity of 
the $K_SK^+K^-$ three-body system is $(-1)^l$, and $\alpha^2$ also gives the 
fraction of CP-even component of the three-body $K_SK^+K^-$ final state. Using
the information from Table~\ref{tab:defitall}, we obtain: 
$\alpha^2 = 0.86\pm0.15\pm0.05$. Note that the $K_SK^+K^-$ three-body final 
state includes the $\phi K_S$ state which is CP-odd. We remove $B^0\to\phi K_S$
events by requiring $|M(K^+K^-)-M_{\phi}|>15$~MeV; the number of remaining 
$K_SK^+K^-$ events is $123\pm14$. The value of the $\alpha^2$  for 
remaining events is: $\alpha_{\rm non~\phi}^2=1.04\pm0.19\pm0.06$. Assuming 
isospin symmetry, we can also use the $B^+\to K^+K^+K^-$ final state instead 
of $B^0\to K^0K^+K^-$ to determine $\alpha^2$. In this case Eq.~\ref{eq:alpha_1}
becomes
\begin{equation}
  \alpha^2 = 
  2\frac{{\cal{B}}(B^+\to K^+K_SK_S)}{{\cal{B}}(B^+\to K^+K^+K^-)} = 
  2\frac{N_{K^+K_SK_S}}{N_{K^+K^+K^-}}\times
  \frac{\varepsilon_{K^+K^+K^-}}{\varepsilon_{K^+K_SK_S}},
  \label{eq:alpha_2}
\end{equation}
which gives $\alpha^2 = 0.82 \pm 0.12 \pm 0.06$ and
$\alpha_{\rm non~\phi}^2 = 0.97\pm0.15\pm0.07$. The two ways of computing
$\alpha^2$ and $\alpha_{\rm non~\phi}^2$ are in good agreement. This is 
evidence for the dominance of the CP-even component in the three-body 
non-$\phi$ $K_SK^+K^-$ final state.

  In conclusion, we have measured branching fractions for charmless
$B$ mesons decays to the three-kaon $K^+K^+K^-$, $K_SK^+K^-$, $K_SK_SK^+$, and
$K_SK_SK_S$ final states. The isospin analysis of the three-kaon final states 
presented here implies several conclusions that should be carefully examined 
further. The three-body $K_SK^+K^-$ final state may  be a good candidate for a
CP violation measurement in $b\to s$ penguin transitions. In this case a
significant increase (by a factor of about four) in the statistics as compared
to the $B^0\to\phi(K^+K^-) K_S$ final state, which is a well known mode 
for the measurement of CP violation in $b\to s$ penguin dominated decays, is
possible.

\section*{Acknowledgments}
We wish to thank the KEKB accelerator group for the excellent
operation of the KEKB accelerator.
We acknowledge support from the Ministry of Education,
Culture, Sports, Science, and Technology of Japan
and the Japan Society for the Promotion of Science;
the Australian Research Council
and the Australian Department of Industry, Science and Resources;
the National Science Foundation of China under contract No.~10175071;
the Department of Science and Technology of India;
the BK21 program of the Ministry of Education of Korea
and the CHEP SRC program of the Korea Science and Engineering Foundation;
the Polish State Committee for Scientific Research
under contract No.~2P03B 17017;
the Ministry of Science and Technology of the Russian Federation;
the Ministry of Education, Science and Sport of the Republic of Slovenia;
the National Science Council and the Ministry of Education of Taiwan;
and the U.S.\ Department of Energy.


\begin{thebibliography}{99}

\bibitem{b2khh}{A.~Garmash {\it et al.} (Belle Collaboration),
         Phys. Rev. D {\bf 65}, 092005 (2002).}
%
\bibitem{bc226}{BELLE-CONF 226, submitted as a contribution paper to
         ICHEP 2002.}
%
\bibitem{KEKB}{ E.~Kikutani ed., KEK Preprint 2001-157 (2001), 
         to appear in Nucl. Instr. and Meth. A.}
%
\bibitem{Belle}{A.~Abashian {\it et al.} (Belle Collaboration),
         Nucl. Instr. and Meth. A {\bf 479}, 117 (2002).}
%
\bibitem{sim}{Events are generated with the CLEO group's QQ program
         (http://www.lns.cornell.edu/ public/CLEO/soft/QQ); the detector
         response is simulated using GEANT, R.Brun {\it et al.},
         GEANT 3.21, CERN Report DD/EE/84-1, 1984.}
%
\bibitem{fisher}{ R.A.~Fisher, Ann. Eugenics {\bf 7}, 179 (1936);
         M.G.~Kendall and A.~Stuart, {\it The Advanced Theory of Statistics},
         2nd ed. (Hafner Publishing, New York, 1968), Vol.III.}
%
\bibitem{vcal}{D.M.~Asner {\it et al.} (CLEO Collaboration), 
         Phys. Rev. {\bf D53}, 1039 (1996).}
%
\bibitem{b2dpi_cleo}{
$\BF (B^+\to \bar{D}^0\pi^+) = (49.7\pm1.2\pm2.9\pm2.2)\times 10^{-4}$ and
$\BF (B^0\to D^-\pi^+)       = (26.8\pm1.2\pm2.4\pm1.2)\times 10^{-4}$, where
the first error is statistical, the second is systematic, and the third is due
to the experimental uncertainty on the production ratio of charged and neutral
$B$ mesons in $\Upsilon(4S)$ decays;
S.~Ahmed {\it et al.} (CLEO Collaboration), hep-ex/0206030.}
%
\bibitem{babar-hhh}{B.~Aubert {\it et al.} (BaBar Collaboration),
         hep-ex/0206004.}
%
\bibitem{CP_phi1}{K.~Abe {\it et al.} (Belle Collaboration),
         Phys. Rev. Lett. {\bf 87}, 091802 (2001).;
         B. Aubert {\it et al.} (BaBar Collaboration), BABAR-PUB-01-03, 2002,
         hep-ex/0201020.}
%
\bibitem{PDG}{D.~E.~Groom {\it et al.} (Particle Data Group), 
         Eur. Phys. J. {\bf C15}, 1 (2000).}
%
\bibitem{Blife}{K.~Abe {\it et al.} (Belle Collaboration), 
         Phys. Rev. Lett. {\bf 88}, 171801 (2002).}

\end{thebibliography}
\end{document}